\documentclass{aa}  %

\def\hi{\relax \ifmmode {\mbox H\,{\scshape i}}\else H\,{\scshape i}\fi}
\def\hii{\relax \ifmmode {\mbox H\,{\scshape ii}}\else H\,{\scshape ii}\fi}
\def\nii{\relax \ifmmode {\mbox N\,{\scshape ii}}\else N\,{\scshape ii}\fi}
\def\sii{\relax \ifmmode {\mbox S\,{\scshape ii}}\else S\,{\scshape ii}\fi}
\def\ha{\relax \ifmmode {\mbox H}\alpha\else H$\alpha$\fi}
\def\me{$^{-1}$}
\def\arcsec{\hbox{$^{\prime\prime}$}}
\def\arcmin{\hbox{$^{\prime}$}}
\def\deg{\hbox{$^{\circ}$}}

\usepackage{graphicx}
\usepackage{float}
\usepackage{epsfig,natbib,txfonts}
\begin{document}
   \title{Where are the  stars of the bar of NGC~1530 forming?}

   \subtitle{}

   \author{A. Zurita\inst{1,2}
          \and I. P\'erez\inst{1,2,3}}

   \offprints{azurita@ugr.es}

 \institute{Depto. de F\'\i sica Te\'orica y del Cosmos, Campus de Fuentenueva,
              Universidad de Granada, 18071-Granada, Spain\\
              \email{azurita@ugr.es}
         \and Instituto Carlos I de F\'\i sica Te\'orica y Computaci\'on
         \and Kapteyn Astronomical Institute, University of Groningen,
              Postbus 800, Groningen 9700 AV, the Netherlands\\
              \email{isa@astro.rug.nl}
             }  

   \date{}

% \abstract{}{}{}{}{} 
% 5 {} token are mandatory
 
  \abstract
  % {The places and parameters which provide suitable conditions for star formation to occur 
  %  in bars are not yet well understood. The strong shear present in the main dust--lanes of 
  %  strong bars have been frequently used to discard any expectation for recent star formation
  %  along strong bars.}
{} 
   {NGC~1530 has one of the strongest bars ever observed and recent star formation sites
    are distributed across its bar. Our aim is to study the photometric properties of the 
    bar and its  \hii\ regions, to elucidate the conditions under which \hii\ regions 
    form and their spatial relation to the principal  dynamical features of the bar.}
   {We  obtained $BVRKs$ and \ha\ photometry of the \hii\ regions of the bar in
    NGC~1530. Broad--band integrated colours and \ha\ equivalent widths were carefully 
    measured and analysed as a function of  position with respect to the  main dust--lanes of 
    the bar.}
  % results heading (mandatory)  through evolutionary synthesis models, and 
   {We measured differences in the \ha\ equivalent widths of the \hii\ regions that are located 
    in the trailing  and leading sides of the bar dust--lane. The possible factors that
    could produce this difference, such as  [\nii] contamination, Lyman--continuum--photon 
    dust--extinction, escape of ionizing radiation, metallicity, IMF and age, were carefully 
    analysed. Age differences were confirmed to be the most plausible explanation. This implies 
    that \hii\ regions that are located  further away from the bar dust--lane in its leading side, 
    downstream from the main bar dust--lane, are older than the rest  by $\sim 1.5-2.5$~Myr. 
    In addition, a clear spatial correlation has been found between: location of \hii\ regions,
    dust spurs on the trailing side of the bar dust--lane, and the loci of maximum 
    velocity gradients parallel to the bar major axis (possibly tracing gas flow towards 
    the main bar dust--lane).These results support the hypothesis 
    that massive stars are forming on the trailing side of the bar dust--lane, and  age as 
    they cross the bar, on a timescale that is compatible with the bar dynamics timescale.}
{} 
\keywords{Galaxies: general --
Galaxies: individual (NGC~1530) --
Galaxies: ISM --
Galaxies: spiral--
Galaxies: kinematics and dynamics --
ISM: HII regions --
}

   \maketitle

%
%________________________________________________________________

\section{Introduction}
\label{intro}
It is commonly believed that the presence or absence of star 
formation (SF) along bars is determined by the bar strength.
Fluid--dynamic simulations of bars (e.g. Athanassoula 1992, 2000) 
support this statement because they predict that the highest gas--density
loci, typically bar dust lanes,  are also the loci of strong shocks and high shear 
in strong bars, which could prevent the collapse of molecular clouds 
and formation of stars.
Therefore, t is frequently stated that SF is not expected to occur within bars of 
strongly--barred galaxies, which is supported by the prototypical 
example NGC~1300 (e.g. Jogee et al. 2002; Tubbs 1982). There are
however several examples of well--known galaxies with strong bars and 
SF along their bars, such as NGC~1530, NGC~1365, 
and NGC~7479, which all have bar strengths $\gtrsim$0.5, as 
measured by Block et al. (2004).

The inhibition of SF due to the high 
velocity of the molecular clouds relative to the bar
was studied numerically  by Tubbs (1982).
It is supported by observations of the spatial anti--correlation between 
the sites of recent SF, which are traced by \hii\ regions, and the sites 
of highest shear along the bar (Zurita et al. 2004; Reynaud 
\& Downes 1998). 
However, there is observational evidence that even if 
SF can be inhibited or suppressed 
along the dust--lanes of strong bars, in some galaxies 
\hii\  regions can be present in other locations of the 
bar structure and not only at the galaxy centres and 
bar ends (e.g.  Martin \& Friedli 1997; Sheth et al. 2002;  
Zurita et al. 2004 and this paper).
What determines the presence of star--forming regions along the bar 
of certain galaxies, and their absence in others?
\hii\ regions are observed in the bars of some galaxies, which implies 
that suitable conditions may exist for SF at locations within the entire
bar structure that are not related to the highest density loci 
predicted by current models of bars.

The extreme physical conditions within bars, which are sites of for example
strong shear, shocks and non-circular velocities 
(e.g. Pence \& Blackman 1984; Athanassoula 1992; 
Reynaud \& Downes 1998; Zurita et al. 2004), 
and significant magnetic--field strength (e.g. Beck 2002),  
make them excellent laboratories for studying the 
physical parameters that, in general, trigger and inhibit star 
formation. This is because there appears to be no obvious 
differences between the \hii\ regions of bars and discs 
(Martin \& Friedli 1999):
their physical properties, dust extinction, and equivalent widths 
are all similar. Observational indications for differences 
between \hii\ regions of bars and  discs have only been found 
in the degree of ionization (Martin \& Friedli 1999), and  the 
\ha\ luminosity function (Rozas et al. 1999), the latter for 
only one galaxy, NGC~7479, which has suffered a minor--merger event 
(Laine \& Heller 1999; Martin, Lelievre, Roy 2000).
\begin{table*}
\caption{Log of the observations relevant to this paper.}
\label{observations}
\begin{tabular}{l l l l l l l l l}     % 7 columns
\hline\hline
                      % To combine 4 columns into a single one
Band& Date & Telescope & Instrument & Filter & Detector & Exp. Time & Seeing\footnote{In final combined image}& Reference\\
\hline
   B & 3 Feb 2006 & 2.56m NOT & ALFOSC &  Bessel B (\#74)  &EEV42-40 2Kx2K &  3$\times$300s   & 0.8\arcsec& (a)\\
   V & 3 Feb 2006 & 2.56m NOT & ALFOSC &  Bessel V (\#75)  &EEV42-40 2Kx2K &  3$\times$300s   & 0.7\arcsec& (a)\\
   R & 3 Feb 2006 & 2.56m NOT & ALFOSC &  Bessel R (\#76)  &EEV42-40 2Kx2K &  4$\times$300s   & 0.8\arcsec& (a)\\
  Ks & 7,8 Oct 2001 & 4.2m WHT  & INGRID &      Ks         &Rockwell HgCdTe 1Kx1K & 2820s   & 1.8\arcsec&   (b)\\
 \ha & 4 Aug 2001   & 1m   JKT  &   JAG  &   Ha6626/44     &SITe2   2Kx2K  &  4$\times$1200s& 1.3\arcsec&   (c)\\
 \hline
\end{tabular}
{\footnotesize \begin{tabular}{l}
$^{1}$ As measured in the final combined image\\
References: $(a)$ This paper; $(b)$ Block et al. (2004); $(c)$ Rela\~no et al. (2005)\\
\end{tabular}}
\end{table*}
   
The places and physical conditions that offer suitable conditions
for star formation inside bars are not yet well understood.
Most of our current knowledge on the location, 
distribution, and properties of star--forming regions in bars
has been derived from a handful of papers centred on their physical properties 
(Martin \& Friedli 1999), and on the general morphology
of the \ha\ emission (e.g. Phillips 1996; Verley et al. 2007b),
and on its relation with the molecular gas (e.g. Sheth et al. 2000, 2002)
or with the stellar bar (Martin \& Friedli 1997; Verley et al. 2007b).

Recent SF has a range of spatial distributions in bars, 
which can be traced by \ha\ emission.
When  \ha\ is detected in a bar, it appears to originate in: (1) \hii\ regions 
distributed along the bar;  (2) the nuclear or
circumnuclear region with little or no emission from the bar, 
or (3) the bar and the nuclear region, i.e. an intermediate 
case between (1) and (2), (Martin \& Friedli 1997; Verley et al. 2007b).
The second distribution appears to be more common in spirals, and its
frequency appears to be rather constant ($\sim$40\%) 
for different Hubble types from Sa to Sc  (Verley et al. 2007b), 
in contrast to the earlier results of Phillips (1996) 
for a smaller sample of galaxies.
Martin \& Friedli (1997) interpreted the different \ha\ bar
morphologies as stages of an evolutionary sequence 
of the bar, which has recently been re--hypothesized by Verley 
et al. (2007b). The sequence 
initiates with SF distributed along the bar. The  gas is then progressively 
transferred from  the bar through  gas inflow towards the centre
of the galaxy until \ha\ emission is only observed in the nuclear 
or circumnuclear region.  This latter stage appears to represent 
most of the bar lifetime, given its observed frequency 
(Verley et al. 2007b).

When \ha\ emission is detected along bars, the major axes of the  
bars defined from the stellar and ionized  gas emissions are 
commonly misaligned by up to $\sim$30\deg (Martin \& Friedli 1997; 
Verley et al. 2007b; Rozas, Zurita, Beckman 2000), 
with the \ha\  bar leading the stellar bar, if we assume trailing spiral arms.
The \ha\ emission along  bars, if present, normally also leads 
the molecular hydrogen, which is traced by CO, or the bar dust--lanes 
by an amount that increases with bar strength and
ranges from 0 to $\sim$800~pc (Sheth et al. 2000, 2002). 

The misalignments between gaseous and stellar components strongly confirm 
that dynamics in bars play an important role in SF. 
The aim of this paper is to study the 
properties of the \hii\ regions of the bar in
NGC~1530  to determine the factors that favour SF 
in the environment of its bar, attending to its dynamical properties. 
NGC~1530 is an excellent target for this study. It is nearby (distance $\sim$37~Mpc), 
isolated (Verley et al. 2007a),
has one of the strongest bars ever observed
(bar strength$\sim$0.73, Block et al. 2004),  and strong 
shear, which has been observed along its prominent straight bar 
dust--lanes (e.g. Reynaud \& Downes 1998; Zurita et al. 2004).  
On the other hand, it presents regions of current SF along the bar, 
in spite of the strong shear and shocks observed, that are not 
limited to the ends of the bar and the nuclear region.

\section{Observations}
\subsection{Broad band optical imaging}
\label{broad_band}
Our optical imaging in the $B, V$ and $R$ bands was obtained in 
service mode at the 2.56m Nordic Optical Telescope (NOT) at the 
Observatorio del Roque de los Muchachos (ORM) in La Palma, with the 
ALFOSC instrument in imaging mode.
The observations were taken on  2006 February 3 in both photometric 
and  excellent seeing conditions, with seeing ranging from 
$\sim$0.6\arcsec up to 0.8\arcsec. The ALFOSC detector is a thinned
2048$\times$2048 E2V  CCD42-40 chip, which provides a field of view 
of $\sim$6.5\arcmin$\times$6.5\arcmin, and a pixel scale of 0.19\arcsec/pix. 
A summary of the observations details, including total integration times, is
given in Table~\ref{observations}.

Three fields of photometric standard stars (Landolt 1992) were taken before 
and after the observations of the galaxy. They covered a wide range in  airmass,
$1.1<X<2.6$, and colours ($0.005<B-R<5.0$).

The data reduction was carried out using standard IRAF\footnote{IRAF 
is distributed by the National Optical Astronomy Observatories, which 
is operated by the Association of Universities for Research in Astronomy, 
Inc. (AURA) under cooperative agreement with the National Science Foundation.}
techniques,  which included overscan subtraction, and bias
and flat field corrections. Cosmic--ray hits were removed from  individual 
images using  the IRAF task  {\tt lacos$\_$im} (van Dokkum 2001).
Afterwards, the sky was subtracted from the individual images, which were then 
aligned and combined to yield one final image for each broad--band filter.

The IRAF packages {\sc apphot} and {\sc photcal} were used to complete aperture 
photometry of the Landolt standard--star fields and derive
the photometric calibration, respectively. The following transformation 
equations were fitted:
\begin{eqnarray*}
B & = & b + b1 + b2 * (B - V) + k_B * X \\
V & = & v + v1 + v2 * (B - V) + k_V * X\\
R & = & r + r1 + r2 * (V - R) + k_R * X 
\end{eqnarray*}
where $B, V, R$ are the Landolt standard--star magnitudes;  $b, v, r$ 
the measured instrumental magnitudes; $X$ is the airmass; 
$b1,v1,r1$ are the zero points per 1 ADU per second for each band;  
$b2,v2,r2$ the coefficients of the colour terms, and $k_B,k_V,k_R$ 
are the extinction constants. The results of the fits are provided 
in Table~\ref{fotometria}. 
The limiting magnitudes for a signal--to--noise ratio per pixel 
equal to 3 are 24.5, 24.2 and 24.4~mag~arcsec$^{-2}$ for $B, V$ and $R$ respectively.
\begin{table}[!h]
\caption{Optical broad--band imaging photometric calibration parameters.}             
\label{fotometria}      
\centering          
\begin{tabular}{c c c c}
\hline\hline       
Band & zero point  &  colour coeff. & extinction coeff.\\ 
     & (for 1 ADU s\me)&                &\\ 
\hline                    
   $B$ & 25.94$\pm$0.01   &  0.029$\pm$0.006 &  -0.20$\pm$0.01\\  
   $V$ & 25.84$\pm$0.01   & -0.062$\pm$0.003 & -0.112$\pm$0.007\\
   $R$ & 25.67$\pm$0.01   &  -0.10$\pm$0.01  & -0.076$\pm$0.007\\
 \hline                  
\end{tabular}
\end{table}

The images of NGC~1530 were astrometrically--calibrated using 
the USNO2 catalogue coordinates for the foreground stars of the galaxy images 
and  the IRAF tasks {\tt ccmap} and {\tt msctpeak}. The accuracy of the astrometric 
calibration was $\sim$0.21-0.29\arcsec.
\begin{figure}[H]
%\resizebox{\textwidth}
\includegraphics[width=0.45\textwidth]{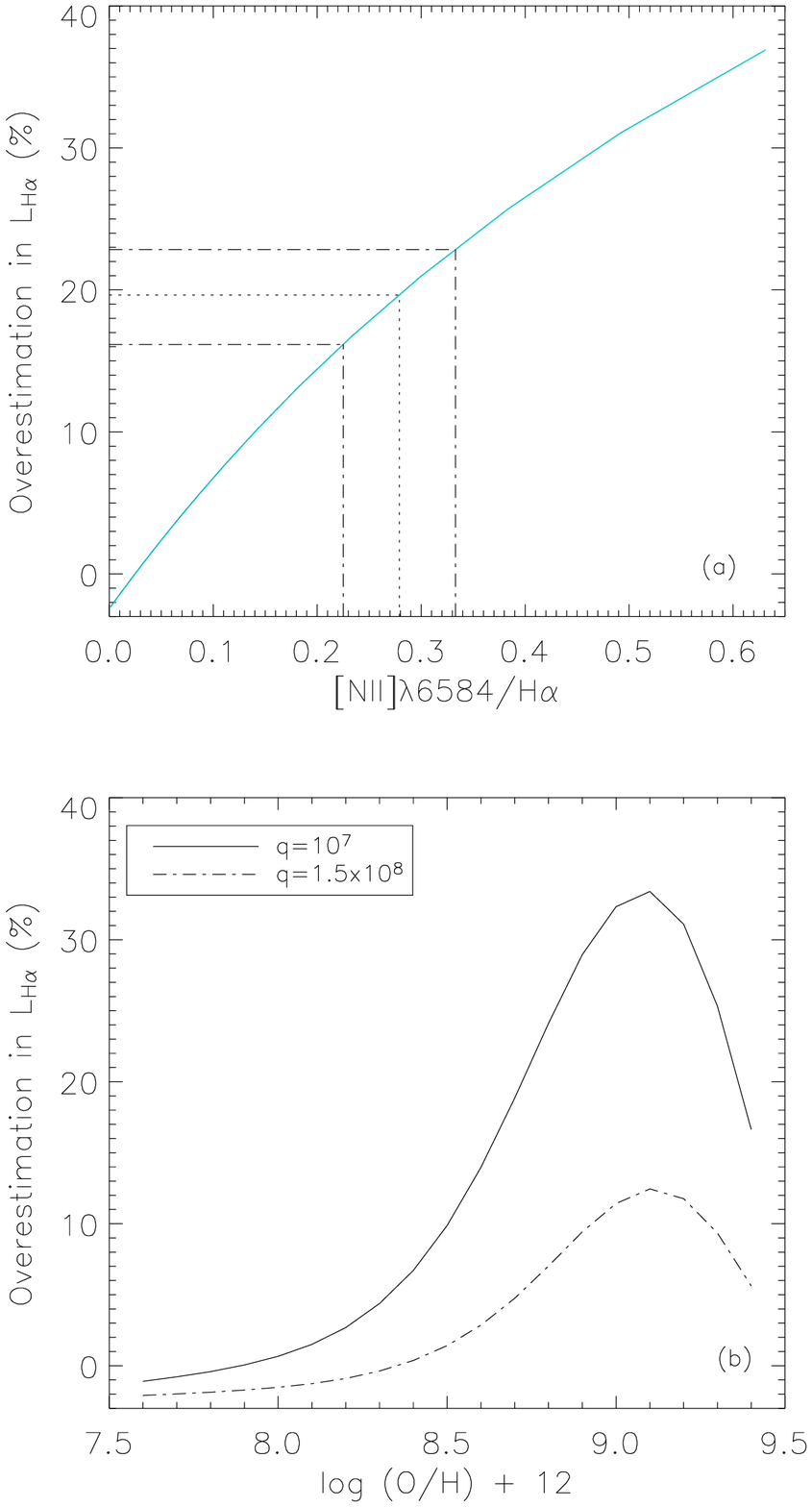}
\caption{{\bf (a)} Percentage of  luminosity that is required to be subtracted from the
measured  \ha\ luminosity to correct for contamination from the [\nii] 
emission lines (at 5648\AA\ and 6584\AA) in the on--line filter, and from the 
inclusion of emission lines in the continuum $R$--band filter, as a 
function of the [\nii]5684/\ha\ ratio.  The dotted line indicates the correction 
applied to the \ha\ luminosity of the bar \hii\ regions of NGC~1530. The dashed lines show 
the corresponding correction for the two \hii\ regions
observed by M\'arquez et al. (2002). {\bf (b)} Same as (a) but as a function of 
the metallicity for two extreme values of the ionization parameter. We have employed 
the [\nii]$\lambda$5684/\ha\  parametrisation  as a function of $q$ and 
metallicity (from  Kewley \& Dopita 2002), and the transmission curve 
of the on and off--line filters for creating the plot. The contribution of 
[\sii]$\lambda\lambda$6717, 6731\AA\ to the continuum flux is negligible. 
Therefore, (a) and (b) can  be used to obtain the overestimation 
in the measured EW$_{\rm{H\alpha}}$ due to the inclusion of the emission
lines mentioned above (see Sect~\ref{observational_effects}).}
\label{contaminacion}
\end{figure}
%Grafico generado con contrib_lineas_plot_paper.pro
% Se ha aplicado esta correccion en los flujos.

Colour maps $B - V$, $B - R$, and $R - V$ were also obtained.
Before creating them, we degraded the spatial resolution of the 
individual $B,V$, and $R$ images to the resolution of the $B$ band image 
(the lowest resolution image), using the  IRAF package 
{\sc daophot}. First, we computed the point spread function (PSF) for 
each image, then we obtained the PSF  matching 
functions  to be convolved with the original $V$ and $R$ images.
Finally,  we produced images of  PSF identical to the PSF of the $B$ band image.
The effective spatial resolution of the optical colour maps was then $\sim$0.8\arcsec.
Some of the colour maps are shown in Fig.~\ref{mosaico_barra} and will be discussed 
in Sect~\ref{morfo}.

\subsection{Broad--band Ks imaging}
NGC~1530 was observed in $K$-short or $K_s$ (2.2$\mu m$) using the INGRID  
camera (Packham et al. 2003), mounted at the Cassegrain focus of the 
4.2m William Herschel Telescope (WHT). 
The details of the observations and the data reduction are given in 
Section 2 of Block et al. (2004). The final reduced $K_s$ image resulted from
a total integration time on source of 2820s under non photometric conditions 
and has a spatial resolution of $\sim$1.8\arcsec\ (FWHM) 
and a field of view of $\sim$4\arcmin$\times$4\arcmin.

As for the optical broadband data, the $Ks$ image was astrometrically calibrated 
using the USNO2 catalogue and following the same procedure as for the
optical data. Approximately 35 
field stars of the image were employed in the fit. The accuracy of the calibration
 was $\sim$0.2\arcsec-0.3\arcsec, and the pixel scale of the detector was calculated 
to be 0.238$\pm$0.001\arcsec/pix.

The image was photometrically calibrated using the Two Micron All Sky Survey (2MASS) 
point--source catalogue using the {\em Aladin} sky atlas (Bonnarel et al. 2000).  Non--saturated 
stars in our image were matched with stars included in the 2MASS 
catalogue, in particular stars brighter than $\sim$15.3~mag. We measured a zero 
point, for 1 ADU in our image, of 25.5$\pm$0.05. Using this calibration and the 
plate scale of the INGRID detector, we estimated that the limiting 
magnitude of the image, at a 3$\sigma$ significance level per pixel, was 
19.9 mag~arcsec$^{-2}$ which is in  good agreement with the value provided  
by  Block et al. (2004), for the same data set, using a different calibration.

\subsection{H$\alpha$ imaging}
\label{ha_imaging}
The \ha\ observations of NGC~1530 were obtained on August 2001 using the CCD 
camera of the 1m Jacobus Kapteyn Telescope (JKT) on the ORM.
The galaxy was observed through a narrow--band (40\AA\ width) \ha\ filter  
for 4800s  and through an $R$ broad--band filter. 
However, for the continuum subtraction our deeper $R$--band image, which was acquired 
using the NOT  (see Sect.~\ref{broad_band}) was employed, after scaling  to match the 
flux level and pixel scale of the original 600s $R$--band image from the JKT.

The observations,  data reduction, flux calibration,
and production of the \hii\ region catalogue are described in detail in
Rela\~no et al. (2005). The continuum subtraction was performed 
as described in Rela\~no et al. (2005) but using the deeper image 
described above.

For the purposes of this paper, the astrometry of the \ha\ and continuum images
was improved  using the USNO2 catalogue and the  procedure described 
above, for the broad--band images of NGC~1530. The accuracy of the calibration
was $\sim$0.24\arcsec. 

%{\bf Contaminacion por las lineas de [NII]}
%***[\nii] emission can affect light in admitted by the bandwidth of the \ha\
%filter at the redshifted wavelengths of the [\nii]$\lambda$6548\AA and [\nii]$\lambda$6584\AA.***
The bandwidth of the  \ha\ filter includes partial emission 
from [\nii] at the corresponding red-shifted wavelengths of the
[\nii]$\lambda$6548\AA\ and [\nii]$\lambda$6584\AA\  emission lines.
In addition, the filter used for continuum subtraction contains 
bright emission lines (namely [\nii]$\lambda$6548, \ha, [\nii]$\lambda$6584 and
[\sii]$\lambda\lambda$6717, 6731). The effects of these contaminations have 
opposite senses. If not taken into account, the former overestimates the emitted \ha, while 
the latter produces an underestimation of  the emitted \ha, 
due to an overestimation of the continuum emission in the \hii\ regions.
\begin{figure*}[!ht]
%\resizebox{\textwidth}
\includegraphics[width=\textwidth]{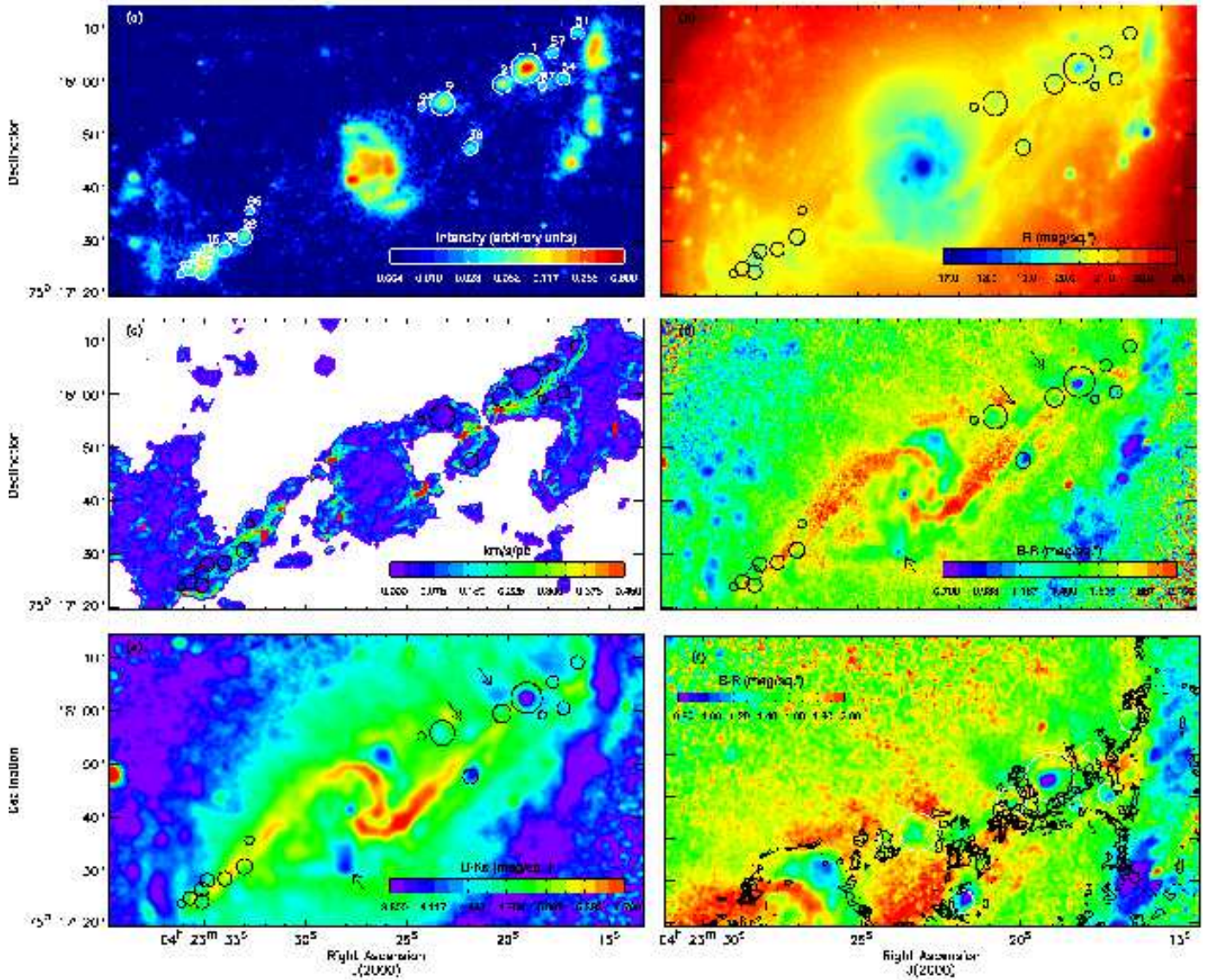}
\caption{ {\bf (a)} \ha\ continuum subtracted image of the bar region of NGC~1530. {\bf (b)}  $R$ band image.  {\bf (c)} Observed projected velocity gradient map in the direction perpendicular to the bar (from Zurita et al. 2004). {\bf (d)} $B-R$ colour map.   {\bf (e)} $B-Ks$ colour map. {\bf (f)} $B-R$ colour map of the NW side of the bar with  overlaid contours of  the velocity gradients in the direction of the bar (velocity gradients from Zurita et al. 2004). We note the spatial correlation between the loci of maximum velocity gradient in the direction parallel to the bar major axis, and the dust spurs in the trailing side of the bar (the information on the velocity gradients is spatially--limited to zones with detectable \ha\ emission). The colour maps were not  corrected for Galactic  extinction. This correction amounts to -0.25 mag/sq\arcsec\ and -0.6 mag/sq\arcsec, for the $B-R$ and $B-Ks$ maps, respectively (Schlegel et al. 1998). Images {\bf (a)} to {\bf (e)} show the same area of the galaxy.  The circles indicate the position and size of the \hii\ regions of the bar. Their identification number, as provided in the catalogue (Rela\~no et al. 2005) is indicated in Fig. 1a. The arrows in panels (d) and (e) indicate blue knots with no counterpart in \ha\ (see Sect.~\ref{morfo} for details.)}
\label{mosaico_barra}
\end{figure*}

A proper correction of the [\nii] contamination requires spectroscopic information 
for  the \hii\ regions. 
Unfortunately, spectroscopy is only available in the literature for two \hii\ 
regions of the bar of NGC~1530 (M\'arquez et al. 2002), both located at 
$~\sim$60-70\arcsec\ from the galaxy centre. 
Greve et al. (1999) completed long--slit spectroscopy 
along the bar of NGC~1530. They  do not, however, report 
the line ratios for integrated fluxes from the \hii\ regions.
The mean \hii\ region line ratios measured by M\'arquez et al. (2002)
were 0.279 and 0.304, for [\nii]$\lambda$6584/\ha\ and [\sii]$\lambda\lambda$6717, 6731/\ha, 
respectively. We  assumed that these 
flux ratios are representative of the \hii\ regions of the bar of NGC~1530.
Using this information, the transmission curves of the \ha\ and $R$--band filter, 
and assuming that [\nii]$\lambda$6584/[\nii]$\lambda$6548=1.34, we estimate
that the [\nii] lines contribute up to $\sim$21.7\% of the measured \ha\ fluxes.
In contrast, the inclusion of emission lines in the continuum filter equals
only 0.02--0.05\% of the total continuum emitted by the \hii\ 
regions, which implies that a correction of $\sim$2\% of the measured \ha\ is required.
%With this information, together with the transmission curves of the \ha\ and $R$--band filter,
%and assuming [\nii]$\lambda$6584/[\nii]$\lambda$6548=1.34 we have estimated that 
%the [\nii] lines contribute up to  $\sim$21.7\%  of the measured \ha\ fluxes, while the inclusion 
%of emission lines in the continuum filter amount to only 0.02--0.05\% of the total 
%continuum emitted by the \hii\ regions (implying a correction of $\sim$2\% on the measured \ha). 
This in turn implies that an average value of 19.7\% of the measured \ha\ luminosity over the 
continuum--subtracted image, needs to be subtracted from the measurements, 
to derive the actual \ha\ emission alone (as can be seen Fig.~\ref{contaminacion}a). 
Variations in the [\nii]/\ha\ ratio are expected to occur from region to region. This 
point will be discussed  further  in Sect.~5.

The \hii\ region \ha\ luminosities given throughout this paper contain the 
calculated  19.7\%  correction, except in the cases in which it is 
specifically noted that measurements come from \ha\ plus [\nii] emission.

\section{Analysis and results}
\label{resultados}

\subsection{Morphology and surface photometry in the bar region}
\label{morfo}

The bar of NGC~1530,  which has a major axis length of $\sim$24~kpc
and a bar class 7 (Buta et al. 2003), is one of the largest
and strongest ever observed, and therefore one of the most extensively 
studied (e.g. Regan et al. 1995, 1996, 1997; 
Downes et al. 1996, Reynaud \& Downes 1998; Greve et al. 1999; Rela\~no 2004;
Zurita et al. 2004).

The bar of NGC~1530  has prominent straight bar dust--lanes 
visible in the colour maps of Fig.~\ref{mosaico_barra}d,e,f, which 
curl around the centre in the inner parts. They are evident in the 
colour maps as areas of high extinction, which are $\sim$0.3 mag and $\sim$0.5 mag 
redder in $R-Ks$ and  $B-Ks$ respectively, than their surrounding regions.
The increasingly red colour along the bar dust--lanes towards the centre 
of the galaxy, could represent an increasing gas density, which is
predicted by numerical models (see e.g. Athanassoula 1992; 
P{\'e}rez et al. 2004).
The southeastern dust--lane is less prominent in the inner part 
than the northwestern one, but the relation is reversed when 
comparing the most external parts of both dust--lanes.
The dust--lanes are offset from the bar 
major axis by $\sim11\deg$.  As shown in Zurita et al. (2004),
and reproduced in Fig.~\ref{mosaico_barra}c of this paper, the 
bar dust--lanes trace strong shocks in the gas flow around the bar.
Other concentrations of dust, which are sometimes called {\em dust spurs}, e.g.
Sheth et al. (2000), are also observed in the colour maps of NGC~1530, and more clearly in the 
$B-R$ maps that have better spatial resolution. These are observed on the trailing
side of the bar\footnote{NGC 1530 rotates clockwise (Zurita et al. 2004).}, 
as red feathers that extend  approximately in a perpendicular direction to  
the main bar dust--lanes (see Fig.~\ref{mosaico_barra}d,f). 
They are more clearly seen in the northwestern side of the bar, 
the closest to the observer, possibly due to projection effects.

\begin{figure}[!h]
%\resizebox{\textwidth}
\includegraphics[width=0.45\textwidth]{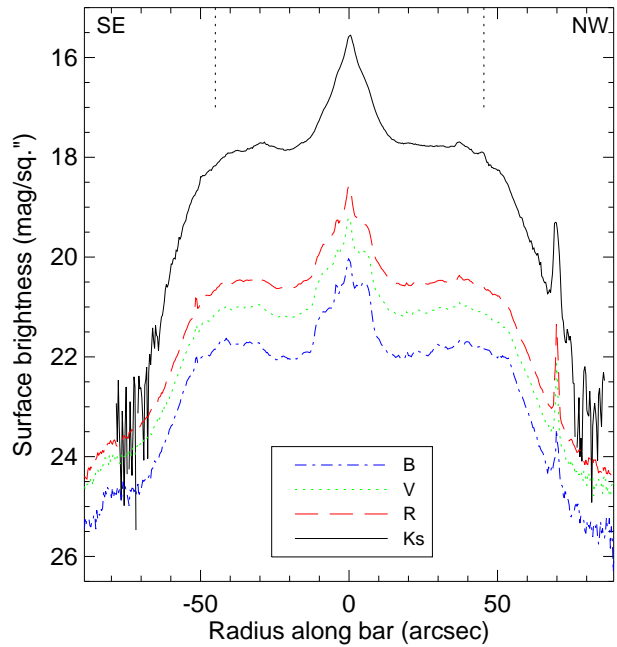}
\caption{Surface brightness profile along the major axis of the bar (PA=121\deg) of 
NGC~1530 for the $B$,$V$,$R$ and $Ks$ bands. The data were corrected for 
Galactic extinction (Schlegel et al. 1998). The dotted line indicates the 
approximate location of the bar end.}
\label{perfiles_along_bar}
\end{figure}

The bar of NGC~1530 emits in \ha\ (see Fig.~\ref{mosaico_barra}a). Most of 
the \ha\ emission originates in the central part of the galaxy or inner spiral.
The remainder of the emission originates in \hii\ regions, which photometric properties
will be further discussed in the next section, plus a contribution from diffuse 
ionized gas (Greve et al. 1999). Its  detection is limited by the 
signal--to--noise ratio of our \ha\ image.
The two brightest \hii\ regions of the bar are located in the northwestern side. 
The regions on this side appear to be  more widely distributed than the ones on the
southeastern side, which are more concentrated in the dust--lane area.
The northwestern side of the bar reveals the presence of \hii\ regions located on
the leading and trailing sides of the bar dust--lanes. As stated in the introduction
of this paper, the presence of \hii\ regions on the trailing side of the bar 
dust--lane is uncommon.  Most studies of recent massive SF in bars, 
have reported the presence of \ha\ emission downstream from
the stellar bar (e.g. Verley et al. 2007b; Martin \& Fiedli 1997; Rozas et al. 2000).
However, in some specific galaxies, such as NGC~6946 (Sheth et al. 2002),  
\ha\ emission has also been detected on the trailing side of the bar.

There is a clear spatial correlation between the location of the \hii\ regions and  
the dust spurs (Figs.~2d,e,f); the former, however, are not necessarily located at the 
end of the spurs, as reported by Sheth et al. (2000) for NGC~5383.
We note also the presence of three blue knots with no 
counterpart in \ha.  These are indicated by arrows in Figs.~\ref{mosaico_barra}d,e.
Two knots are located on the trailing side 
of the bar dust--lane, towards the north and south--east of region 
number 2, and a third one, at 
approximately 15\arcsec\ south of the galaxy centre, 
close to a CO clump (Reynaud \& Downes 1998).
It would be interesting to  investigate their origin, which may be linked to 
 early phases of massive star--formation complexes.

\begin{figure*}[ht!]
%\resizebox{\textwidth}
\includegraphics[width=\textwidth]{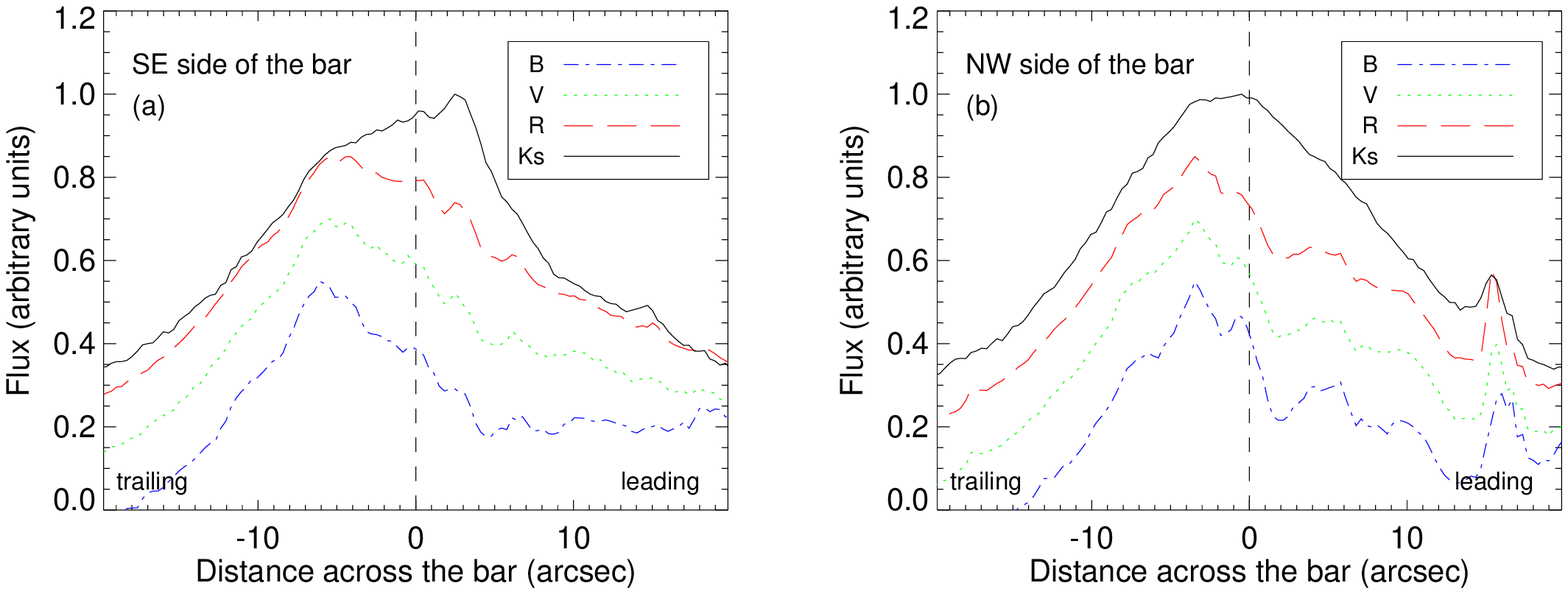}
\caption{Bar minor--axis profiles for the $B$, $V$, $R$ and $Ks$ broad bands 
for the SE (a) and NW (b) half sides of the bar of NGC~1530. The dashed vertical 
line indicates the location of the bar major axis. Negative and positive
distances correspond to the trailing and leading sides of the bar respectively. 
Each profile was normalized to its maximum value and afterwards  the  
$Ks$, $R$, $V$ and $B$ band profiles were displaced  by 0, -0.15, -0.30 
and -0.40 arbitrary units respectively, to facilitate inspection.}
\label{perfiles_flujo}
%\end{figure*}
%\begin{figure*}[!h]
%\resizebox{\textwidth}
\vspace{0.2cm}
\includegraphics[width=\textwidth]{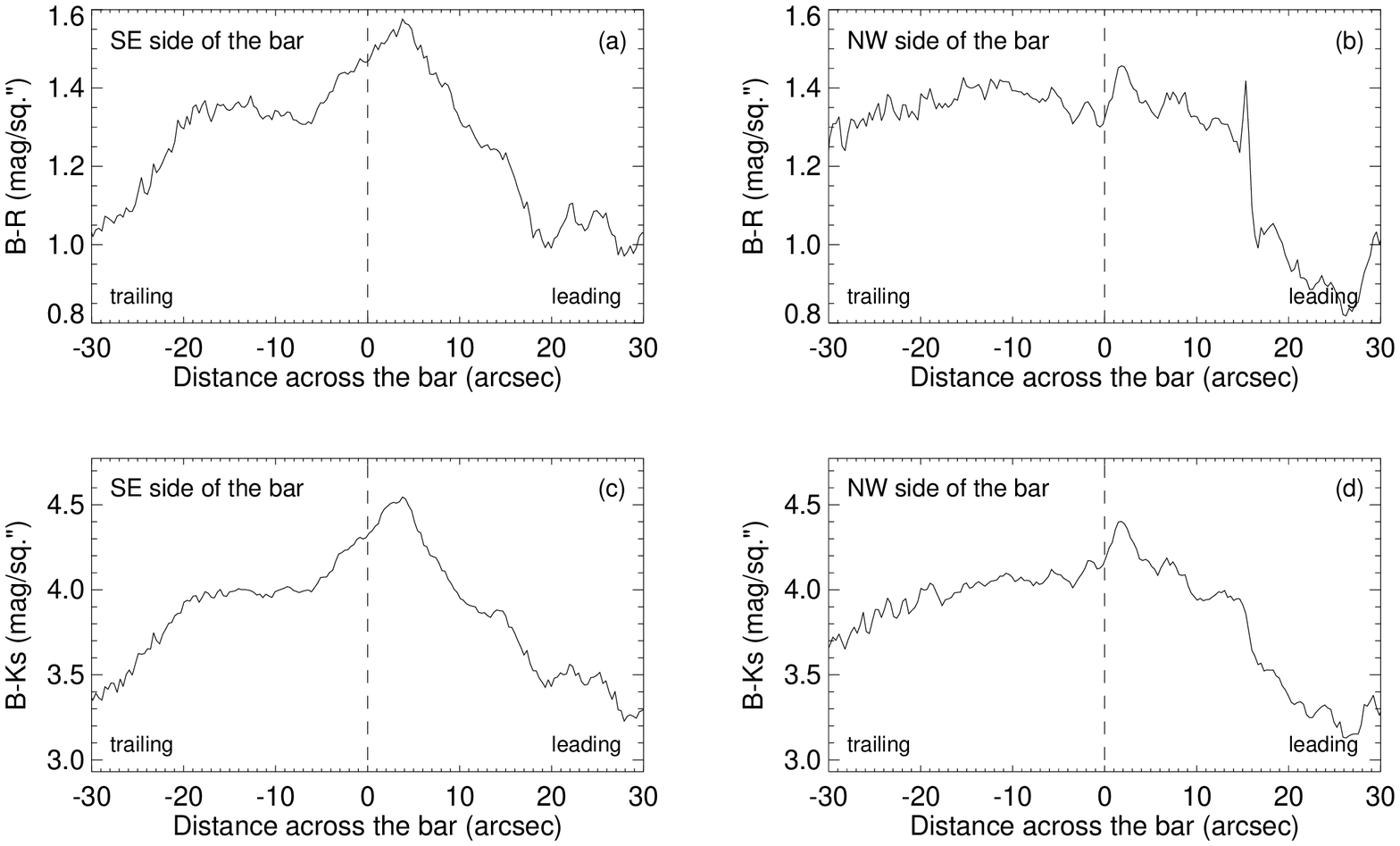}
\caption{$B-R$ (top) and $B-Ks$ (bottom) colour profiles across the bar of NGC~1530 for 
both semi-bar halves (SE on the left panels, (a) and (c), and NW on the right panels, (b) and (d)). As 
in Fig.~\ref{perfiles_flujo}, the dashed vertical line indicates the bar major--axis position,
and  negative distances correspond to the bar trailing side. We note the steeper 
colour gradient of the bar leading side. The dust--lanes appear in the plots 
as  local (redder) peaks at $\sim2-4$\arcsec\ from the bar major axis on the 
leading side. The profiles were corrected for Galactic extinction (Schlegel et al. 1998).}
\label{perfiles_color}
\end{figure*}

To study the light distribution of the NGC~1530 bar, 
broad--band surface brightness profiles were extracted both along and across
the bar, as shown in  Figs.~\ref{perfiles_along_bar} and ~\ref{perfiles_flujo}, 
respectively. They were derived  from broad--band maps 
in which the  star--forming regions of the bar were masked out; we provide more 
details on the masking procedure in  Sect.~\ref{metodos_fondos}.
Figure~\ref{perfiles_along_bar} shows $B$, $R$, $V$, and $Ks$ surface brightness 
profiles  along the bar of NGC~1530. 
The profiles were obtained by averaging flux in slices of  $\sim$23\arcsec\ length,
which were orientated at a  perpendicular angle to the bar major axis. This slice 
length choice matches  the {\em width} of the bar approximately,
because at $\sim$11\arcsec\ from the bar major axis in the direction 
perpendicular to the bar, the bar peak flux has decreased down to $\sim50$\%.
The figure clearly shows flat profiles along the bar region;
we indicate its approximate limits using dotted lines. According to  Elmegreen 
\&  Elmegreen (1985), this behaviour is expected for NGC~1530 since they found that 
Hubble--type galaxies earlier than Sb--Sbc, tend to have uniform intensities along 
the bar length. However, Seigar \& James (1998), for a larger sample, found that 
only $\sim$60\% of bars were flat in early--type galaxies.

Averaged surface--brightness profiles in the direction perpendicular to the bar (cross--sections) are 
shown in Fig.~\ref{perfiles_flujo}.  The profiles were obtained by averaging flux in slices of  
$\sim$17\arcsec\ length orientated in the direction of the bar major axis and centred roughly 
midway between the galaxy centre and the bar ends. This slice length is of the maximum value that 
avoids the inclusion of flux from the NGC 1530 inner spiral and from enhanced emission 
at the bar ends. The same procedure was completed on both halves of the bar. 
The southeastern  bar side profile is plotted in Fig.~\ref{perfiles_flujo}a and the 
northwestern bar side in Fig.~\ref{perfiles_flujo}b. The profiles are asymmetric. 
Within $\sim\pm10\arcsec$ from the bar major axis, as represented by a vertical 
dashed line, the profiles are  clearly steeper in the trailing side than 
in the leading side in the optical bands, but this difference appears to become opposite 
in the $Ks$--band profile. Asymmetric cross--sections, with steep leading edges 
were reported by Seigar \& James (1998) using $K-$band photometry.
Galaxy types earlier than Sb appear to show symmetric cross--sections (Ohta et al. 1990).
This asymmetry is also present in the colour profiles in Fig.~\ref{perfiles_color},
which show a far steeper colour gradient for the leading side of the bar.
Although a major dust concentration on the trailing side of the bar could 
 contribute to the observed asymmetry, the major contribution must come from 
matter constituting the bar, which is evident in the broad--band images 
of the bar (e.g.  Fig.~\ref{mosaico_barra}b).

\subsection{\ha\ equivalent widths and BVRKs photometry 
of the \hii\ regions of the bar}
The \hii\ region catalogue of NGC~1530 was presented in Rela\~no 
et al. (2005).  A total of 119 \hii\ regions were catalogued (excluding the
centre of the galaxy), with 
\ha\ luminosities ranging from $\sim6.3\times10^{37}$erg~s\me\ to 
$\sim 10^{40}$erg~s\me. Of these, approximately $\sim$20 \hii\ regions 
are located on the bar of the galaxy.

To investigate the properties of the \hii\ regions of the 
bar of NGC~1530, we selected the  17 brightest catalogued \hii\ regions 
of the bar of the galaxy; the luminosity of three of these regions  was too close to 
the detection limits and they were rejected. For the remaining regions, we measured 
the H$\alpha$ equivalent widths (EW$_{\rm{H\alpha}}$) and their broad--band emission.

The \ha\ fluxes were obtained from the catalogue. 
The fluxes in $B$, $V$, $R$ and $Ks$ bands were measured using aperture 
photometry with GAIA\footnote{GAIA is a derivative of the Skycat catalogue 
and image display tool, developed as part of the VLT project at ESO. Skycat 
and GAIA are free software under the terms of the GNU copyright.}.
Circular or elliptical apertures 
were employed, with radii or axes defining integration areas 
compatible with the area of the \hii\ regions as catalogued from the 
\ha\ image and the REGION software (Rela\~no et al. 2005). 
The same apertures were employed for all broad--band images, after a proper 
alignment and re-sampling of  the FITS images to match the \ha\ image
pixel size (0.33\arcsec).
In a few cases in which the overlapping of \hii\ regions was important, the area 
of integration was slightly modified with respect to the one given in 
the \hii\ region catalogue, and  we measured the \ha\ emission 
within the new apertures. The measured differences between the \ha\ luminosities 
obtained with REGION and GAIA were always smaller than the errors in the 
\ha\ luminosities. 

When measuring broad--band fluxes from \hii\ regions, we were interested only 
in the radiation emitted by the ionizing cluster of the nebula. 
However, within the integration aperture we are measuring the contribution from both 
the ionizing cluster and the underlying population of the disc/bar 
of the galaxy. 
To derive the flux from the cluster  alone, it was therefore 
necessary to subtract the contribution from the disc/bar. Subtracting the 
contribution from the bar/disc is not, however, straightforward because the
light distribution of the galaxy is not smooth, which can be seen in 
Fig.~\ref{mosaico_barra}b. This subtraction is the major source of uncertainty
in the measured \hii\ region broad--band fluxes and equivalent widths.

We followed four different approaches to quantify the contribution of the 
underlying population. All assume that \hii\ regions are transparent 
to the radiation emitted by this population. Each method is described in 
detail below.

\subsubsection{Methods for quantifying the local background contribution}
\label{metodos_fondos}

\begin{enumerate}
\item{\bf Annular apertures around the \hii\ regions.}
%
% Ver: /home/azurita/astro/ews/NGC1530/cala_regiones.pro
%
The local background for each broad  band was determined, using the corresponding 
image, to be the median value within an annular aperture around each \hii\ 
region. The median was calculated after a number of iterations using a 
sigma--clipping algorithm, which rejects pixels deviating by more than  
twice the standard deviation from the median value.
The inner radius of the aperture was equal to the {\em effective} radius of 
the \hii\ region. An annular width of 5 pixels ($\sim$1.7\arcsec) was used 
in all cases, which corresponds to $\sim$ 270~pc, and provides a good compromise 
between good sampling and ensuring that the measurement corresponds to the
neighbourhood of the \hii\ region. 
The areas covered by catalogued \hii\ regions were masked out before 
computing the median,  to avoid contamination from continuum emission 
of nearby \hii\ regions.

%For avoiding contamination from continuum emission from nearby \hii\ regions
%the areas covered by catalogued \hii\ regions were masked out before %
%computing the median.
%This is a procedure similar to the one described in Zurita et al. (2001) 
%for estimating the contribution of the diffuse ionized gas in the areas 
%of the galactic disc covered by the \hii\ regions.

\item{\bf Growth curves.}
%
% Ver: /home/azurita/astro/ews/NGC1530/curva_crecimiento.pro
%
In most \hii\ regions, the continuum emission is less extended 
than the \ha\ emitting area, and useful information for determining the local 
background can be obtained from within the \hii\ region area, which was masked 
out in the previous method,  determined from the \ha\ image in the  
\hii\ region catalogue. We derived azimuthally--averaged radial 
surface--brightness  profiles for each \hii\ region for each broad--band.
Then, for  each one we determined a suitable radial range not contaminated by
emission from the ionizing cluster,  which could be used to determine
the local background. 
Finally, we plotted and inspected the growth curve or cumulative flux, as a function 
of the aperture radius, after subtracting the local background value determined previously.
A few iterations were often required before convergent results were obtained.
This procedure was repeated interactively for each bar \hii\ region in each band.

\item{\bf Light profiles perpendicular to the bar.}
%
% Ver: /home/azurita/astro/ews/NGC1530/perfiles.pro
%
Surface--brightness profiles along the direction perpendicular to the bar were 
obtained  at the location of each \hii\ region in each broad--band image. 
The value at each pixel of the profile was obtained by median--averaging, in 
the direction of the bar, the flux  for a number of pixels 
approximately equal to the  \hii\ region size.  From these initial 
profiles a set of smoothed profiles were 
created applying a median filter  with smoothing window sizes  between 3 and 
7 times the \hii\ region diameter. Afterwards, these profiles were interactively 
inspected and compared, and the  profile  of the highest quality was selected
using the criteria that the profiles did not contain detectable continuum emission
from the ionizing clusters of the  \hii\ region under consideration, and had not
averaged--out  excessively large--scale brightness fluctuations in the bar.

\item{\bf Average light profiles perpendicular to the bar.}
A fourth estimate of the local background was obtained using two integrated 
surface--brightness profiles in the direction perpendicular to the bar. 
These were obtained as described above but, the value at each pixel of each 
profile was calculated instead by median--averaging along each side of the bar within regions 
of length $\sim$20 and $\sim$30\arcsec\ for the eastern and western sides of 
the bar, respectively, which avoided the galaxy inner region and the edges of the bar.
The local background for each region was then obtained by interpolation on the 
corresponding (eastern or western) averaged, smoothed profile at the position 
across the bar of the \hii\ region.
\end{enumerate}

\begin{table}[!h]
\caption{Averaged local background values for the \hii\ regions of the bar. Values were 
corrected for Galactic extinction (Schlegel et al. 1998).}             
\label{fondos_locales_tabla}      
\centering          
\begin{tabular}{r c c c  c }
\hline\hline       
                      % To combine 4 columns into a single one 
ID & B  &  V & R   & K          \\ 
   & mag/sq\arcsec &mag/sq\arcsec & mag/sq\arcsec & mag/sq\arcsec    \\ 
\hline                    
   9  &  21.90 $\pm$0.04 & 21.06$\pm$0.02 & 20.48 $\pm$0.02 & 17.64 $\pm$ 0.04   \\ 
   1  &  21.68 $\pm$0.03 & 20.95$\pm$0.03 & 20.40 $\pm$0.02 & 17.63 $\pm$ 0.04   \\ 
  21  &  21.82 $\pm$0.04 & 21.06$\pm$0.03 & 20.43 $\pm$0.02 & 17.58 $\pm$ 0.04   \\ 
  94  &  21.87 $\pm$0.05 & 21.03$\pm$0.03 & 20.40 $\pm$0.02 & 17.56 $\pm$ 0.04   \\ 
  57  &  21.63 $\pm$0.04 & 20.87$\pm$0.03 & 20.33 $\pm$0.02 & 17.63 $\pm$ 0.04   \\ 
  36  &  22.05 $\pm$0.05 & 21.23$\pm$0.02 & 20.65 $\pm$0.02 & 17.75 $\pm$ 0.04   \\ 
  54  &  21.83 $\pm$0.04 & 21.09$\pm$0.03 & 20.50 $\pm$0.02 & 17.73 $\pm$ 0.04   \\ 
  51  &  21.73 $\pm$0.03 & 21.00$\pm$0.03 & 20.41 $\pm$0.02 & 17.65 $\pm$ 0.04   \\ 
  87  &  21.73 $\pm$0.05 & 20.99$\pm$0.03 & 20.44 $\pm$0.02 & 17.65 $\pm$ 0.05   \\ 
  86  &  22.10 $\pm$0.04 & 21.27$\pm$0.02 & 20.65 $\pm$0.02 & 17.78 $\pm$ 0.04   \\ 
  28  &  21.88 $\pm$0.04 & 21.04$\pm$0.03 & 20.41 $\pm$0.02 & 17.50 $\pm$ 0.06   \\ 
  38  &  21.82 $\pm$0.04 & 20.99$\pm$0.03 & 20.33 $\pm$0.02 & 17.49 $\pm$ 0.06   \\ 
  16  &  21.70 $\pm$0.04 & 20.94$\pm$0.03 & 20.41 $\pm$0.02 & 17.62 $\pm$ 0.05   \\ 
  13  &  21.71 $\pm$0.04 & 20.93$\pm$0.03 & 20.37 $\pm$0.03 & 17.59 $\pm$ 0.05   \\ 
  12  &  21.73 $\pm$0.04 & 20.98$\pm$0.03 & 20.38 $\pm$0.02 & 17.57 $\pm$ 0.06   \\ 
  92  &  21.80 $\pm$0.05 & 21.08$\pm$0.04 & 20.48 $\pm$0.03 & 17.71 $\pm$ 0.06   \\ 
  37  &  21.78 $\pm$0.05 & 21.09$\pm$0.05 & 20.52 $\pm$0.03 & 17.71 $\pm$ 0.06   \\ 
\hline                  
\end{tabular}
\end{table}

\subsubsection{Analysis of measured local background values}
Using the four previously--derived local--background estimates for each \hii\ region 
in each broad band (Sect.~\ref{metodos_fondos}), we computed the average 
value  (see Table~\ref{fondos_locales_tabla}).
Only methods that yielded  realistic values of the underlying stellar--population 
emission were considered when computing the average. This average value 
was corrected afterwards  for Galactic extinction (Schlegel et al. 1998). The errors 
were obtained by adding in quadrature  errors from each individual 
method.
These, in turn, include the uncertainties in the zero--point photometric calibration, and
the extinction constant, and the error inherent to each method.

\begin{figure*}[!ht]
%\resizebox{\textwidth}
\includegraphics[width=\textwidth]{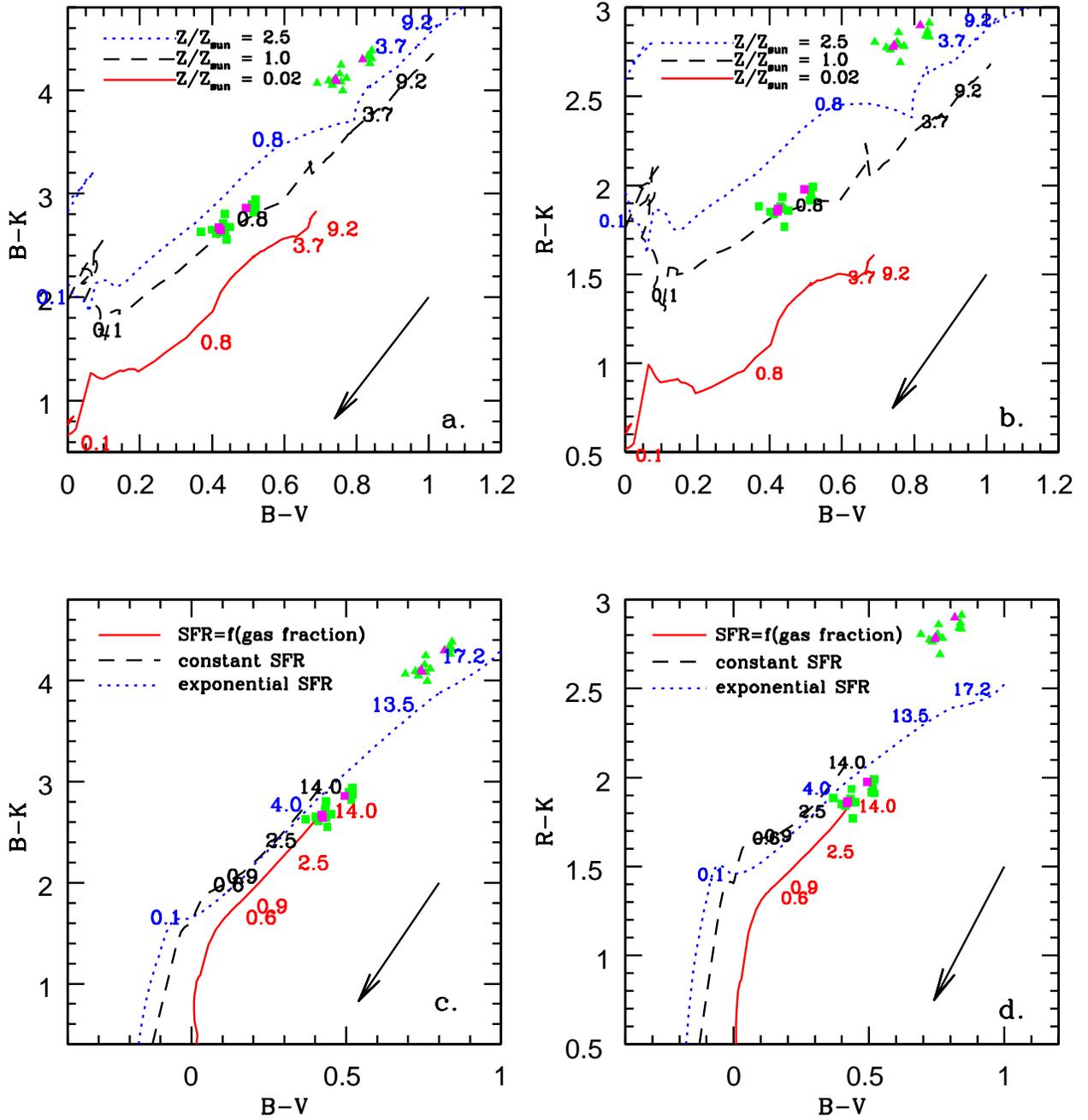}
\caption{ {\bf (a)} and {\bf (b)}: colour--colour plots for single stellar--population models 
(Bruzual \& Charlot  2003) for different metallicities: Z/Z${_\odot}$=0.02 (solid line), 1.0 
(dashed), 2.5 (dotted).  {\bf (c)} and {\bf (d)}: colour--colour plots for stellar population synthesis
models assuming solar metallicity and a different star formation history:  a star formation rate (SFR) 
decreasing with time through a  dependence on the galaxy gas mass fraction (solid line) 
($SFR=0.3\times10^{-3}$M$_\odot$ Myr$^{-1}\times$(M$_{gas}/$M$_{total}$); Fioc \& Rocca--Volmerange 1997);  
a constant rate of star formation (dashed line) ($SFR=0.5\times10^{-4}$M$_\odot$ Myr$^{-1}$; 
Fioc \& Rocca--Volmerange 1997); and an exponentially decreasing with time SFR  (dotted line)
($SFR = 0.25\times10^{-3}$M$_\odot$ Myr$^{-1}\times \exp{(- t / 4\, Gyr)}$; Bruzual \& Charlot  2003).
The numbers along the tracks indicate the stellar population age (in Gyr) at each position.
Filled squares and triangles represent our measurements of local background values 
after and before extinction correction, respectively. Filled magenta symbols indicate local 
background determinations for the \hii\ regions located furthest away from the bar dust 
lane on the leading side, while green filled symbols correspond to 
the rest of the measurements. The extinction correction includes Galactic extinction (Schlegel 
et al. 1998), and extinction implied by an empirically determined colour excess E($B-V$)=0.3 and  
Calzetti et al. (2000) attenuation curve, which gives the following extinction in 
magnitudes for each band: A$_B$=1.6 mag, A$_V$=1.2 mag, A$_R$=1.0 mag, A$_{Ks}$=0.1 mag. 
The arrow in each plot indicates the extinction correction 
corresponding to A$_V$=1 mag.}
\label{fondos_locales}
\end{figure*}

To determine how accurately our local--background measurements represent 
the underlying disc/bar stellar population, we compared the colours with 
predictions of stellar population--synthesis models (Bruzual \& Charlot  2003;  
Fioc \& Rocca--Volmerange 1997) of different metallicities and SF histories.
%In order to test how well our  obtained local background values represent the underlaying 
%disc/bar stellar population, we have compared their colours with the ones predicted by stellar 
%population synthesis models (Bruzual \& Charlot  2003;  Fioc \& Rocca--Volmerange 1997) for different 
%metallicities and SF histories. 
Some of these comparisons are shown in  Fig.~\ref{fondos_locales}. The measured
local--background colours are compatible with  a   population  of  
age $\sim$0.7--1~Gyr when a single burst of SF is assumed, or $\sim$4--18~Gyr for continuous SF.  
In all cases when assuming the Calzetti et al. (2000) attenuation curve,
we must assume a colour excess for our observational data, after 
Galactic extinction correction, in the range E($B-V$)=0.25--0.35\footnote{A similar colour  
excess is required when the Cardelli et al. (1989) parametrization of the interstellar extinction
is employed for $R_V\sim4$.} to obtain colours that are
compatible with the models for a wide range of metallicities and star--formation rates. 
It is beyond the scope of this paper to characterize the stellar population of the NGC~1530 
disc/bar. 
%Since the measured local--background colours agree well with model--galaxy colours, 
%we can however be confident that we are ***robustly determining*** the contribution of the disc/bar 
%to the  \hii\ region fluxes.
However, the fact that the measured local--background  colours agree well 
with model--galaxy colours, provides  confidence on our determination of the 
underlying stellar--population contribution to the \hii\ region fluxes. 
Furthermore,  the need for a similar 
extinction correction in all SF scenarios  indicates that a colour excess in the range 
E($B-V$)=0.25--0.35  could represent the average extinction in the bar zone. 
The derived value implies a higher extinction, by approximately $0.6$~mag in the $B$ band, in the 
NGC 1530 bar zone, than disc--averaged extinction values for galaxies of the same morphological 
type  (Bell \& Kennicutt 2001); but it is however in 
good agreement with extinction measurements of opaque areas of  galaxies, such as spiral arms 
 (Calzetti 2001), which is expected given the prominence of the bar dust--lanes in 
 NGC 1530. The spread in colours of the individual data points in Fig.~\ref{fondos_locales} 
could represent  differential dust extinction between different local--background regions. 
Under that assumption, the observed spread in colours implies a standard deviation in visual 
extinction between local--background regions of $A_V\sim0.2$ mag.

\subsubsection{Results on \hii\ regions photometry}
\label{results_hii}
For each \hii\  region of the bar, four different estimates of the flux in each 
broad--band image were obtained, which each corresponded to a different estimate 
of the local background as described in Sect.~\ref{metodos_fondos}. Unique values of the 
EW$_{\rm{H\alpha}}$ and the $B, V, R$, and $Ks$  magnitudes were then 
calculated for each \hii\ region.
 These were measured to be the mean values of the maximum and minimum of the 
\hii\ region fluxes that deviated by less than twice the standard deviation from 
all available flux estimates.
The uncertainties in the 
EW$_{\rm{H\alpha}}$ and the broad--band magnitudes were computed using the difference 
between the maximum and minimum values. They therefore represented the range of values 
covered when using the different estimates of the local background, which are 
the primary  source of uncertainty and much larger than the errors due to  
photometric calibration or  signal--to--noise limitations.
The results are shown in Table~\ref{results_HII}. 
%***********Obtencion de fondos locales promedio:  all_backs.pro
\begin{table*}
\caption{Column 1: Identification number of the \hii\ regions (from the 
NGC~1530 \hii\ region catalogue, Rela\~no et al.~2005); column 2: decimal 
logarithm of the \ha\ luminosity (in erg s\me). The errors represent 
uncertainties in the \ha\ luminosity due to the image noise and the 
uncertainties in the determination of the \hii\ region area 
(as described in Zurita 2001).; column 3: decimal logarithm 
of the \ha\ equivalent width of the \hii\ regions; columns 4 to 7: $B, V, R, 
Ks$ broad band magnitudes of the \hii\ regions corrected for Galactic 
extinction (Schlegel et al. 1998); columns 8 to 11: number of 
data points used for the determination of the broad band colour in the 
corresponding band (see Sect.~\ref{results_hii} for more details). The 
errors in the broad band magnitudes include  uncertainties due to the 
determination of the local background for each \hii\ region (see 
Sect.~\ref{metodos_fondos} for details), which  dominate the error bars. The 
asterisks highlight the \hii\ regions data which rely on only one measurement of
the local background, and which are therefore more uncertain.}
\label{results_HII}      
\centering          
\begin{tabular}{r c c c  c c c c c c c }
\hline\hline       
                      % To combine 4 columns into a single one 
Region & log L$_{\rm{H\alpha}+[N II]}$  &  log EW$_{\rm{H\alpha}+[N II]}$ & $B$ & $V$ & $R$   & $Ks$   &  N$_B$  &  N$_V$  &  N$_R$  &  N$_K$ \\ 
  ID   &  (in erg s\me)     &   (in \AA)  & (mag)    & (mag)    & (mag)      & (mag)       &         &         &         &         \\ 

\hline 
   9  &  39.45$\pm$0.03  &  2.78 $\pm$  0.07   & 20.2 $\pm$  0.3   & 19.9 $\pm$  0.1   & 19.1 $\pm$  0.2   & 16.8 $\pm$  0.4   &  4  &  4  & 4  &  4\\
   1  &  40.01$\pm$0.01  &  2.79 $\pm$  0.07   & 18.4 $\pm$  0.2   & 18.3 $\pm$  0.2   & 17.7 $\pm$  0.2   & 15.5 $\pm$  0.3   &  4  &  4  & 4  &  4\\
  21  &  39.23$\pm$0.05  &  2.8  $\pm$  0.2    & 21.1 $\pm$  0.5   & 20.4 $\pm$  0.6   & 19.8 $\pm$  0.5   & 17.3 $\pm$  0.3   &  4  &  4  & 4  &  4\\
  94  &  38.3$\pm$0.2    &  3.1  $\pm$  0.3    & 23.2 $\pm$  0.3   & 23.3 $\pm$  0.6   & 22.8 $\pm$  0.7   & 19.8 $\pm$  0.7   &  4  &  4  & 4  &  2\\
  57  &  38.71$\pm$0.09  &  3.0  $\pm$  0.3    & 22.0 $\pm$  0.5   & 21.5 $\pm$  0.4   & 21.2 $\pm$  0.7   & 18.6 $\pm$  0.4   &  4  &  3  & 4  &  4\\
  36  &  38.91$\pm$0.06  &  2.35 $\pm$  0.05   & 19.8 $\pm$  0.1   & 19.7 $\pm$  0.1   & 19.4 $\pm$  0.1   & 19 $\pm$  1     &  4  &  4  & 4  &  4\\
  54  &  38.74$\pm$0.08  &  2.40 $\pm$  0.07   & 20.7 $\pm$  0.2   & 20.3 $\pm$  0.1   & 19.9 $\pm$  0.2   & 18.0 $\pm$  0.3   &  4  &  4  & 4  &  4\\
  51  &  38.78$\pm$0.08  &  2.7  $\pm$  0.1    & 22.1 $\pm$  0.8   & 21.4 $\pm$  0.7   & 20.9 $\pm$  0.5   & 20 $\pm$  2 $^{*}$   &  4  &  4  & 4  &1\\
  87  &  38.3$\pm$0.1    &  2.26 $\pm$  0.09   & 21.5 $\pm$  0.3   & 21.1 $\pm$  0.3   & 20.6 $\pm$  0.2   & 18.4 $\pm$  0.1   &  4  &  4  & 4  &  4\\
  86  &  38.4$\pm$0.2    &  3.15 $\pm$  0.04   & 23.6 $\pm$  0.1   & 23.6 $\pm$  0.1   & 22.6 $\pm$  0.1   & 19.0 $\pm$  1.0   &  2  &  2  & 2  &  2\\
  28  &  39.01$\pm$0.05  &  2.83 $\pm$  0.07   & 22.2 $\pm$  0.8   & 21.3 $\pm$  0.5   & 20.3 $\pm$  0.2   & 17.5 $\pm$  0.8   &  4  &  4  & 4  &  4\\
  38  &  38.87$\pm$0.07  &  3.4  $\pm$  0.3    &          --       &        --         & 21.9$\pm$0.7 $^{*}$ & 17.5 $\pm$  0.2   &  0  &  0  & 1  &  4\\
  16  &  39.35$\pm$0.03  &  3.0  $\pm$  0.2    & 21.0 $\pm$  0.3   & 20.5 $\pm$  0.4   & 19.8 $\pm$  0.5   & 18.3 $\pm$  0.9   &  4  &  4  & 4  &  3\\
  13  &  39.43$\pm$0.03  &  2.77 $\pm$  0.08   & 20.2 $\pm$  0.2   & 19.9 $\pm$  0.2   & 19.1 $\pm$  0.2   & 16.6 $\pm$  0.1   &  3  &  3  & 3  &  3\\
  12  &  39.43$\pm$0.03  &  2.69 $\pm$  0.08   & 20.2 $\pm$  0.2   & 19.5 $\pm$  0.2   & 18.9 $\pm$  0.2   & 16.1 $\pm$  0.1   &  3  &  3  & 3  &  3\\
  92  &  38.3$\pm$0.2    &  3.1  $\pm$  0.2    & 23.0 $\pm$  0.3   & 22.5 $\pm$  0.2   & 22.8 $\pm$  0.6   &  --               &  4  &  3  & 3  &  0\\
  37  &  38.88$\pm$0.07  &  2.9  $\pm$  0.1    & 23 $\pm$  1 $^{*}$    & 21.3 $\pm$ 0.5 $^{*}$  & 20.5 $\pm$  0.2   & 17.5 $\pm$ 0.4  & 1 & 1 & 3&4\\
 \hline                  
\end{tabular}
\end{table*}

For some \hii\ regions, one or several   estimates of the local--background 
implied a negative value for the flux of the ionizing cluster. 
Those measurements were rejected, decreasing the 
number of values from which the final broad--band fluxes were computed 
(columns 8 to 11 of Table~\ref{results_HII}).

\begin{figure*}[!ht]
%\resizebox{\textwidth}
\includegraphics[width=1.0\textwidth]{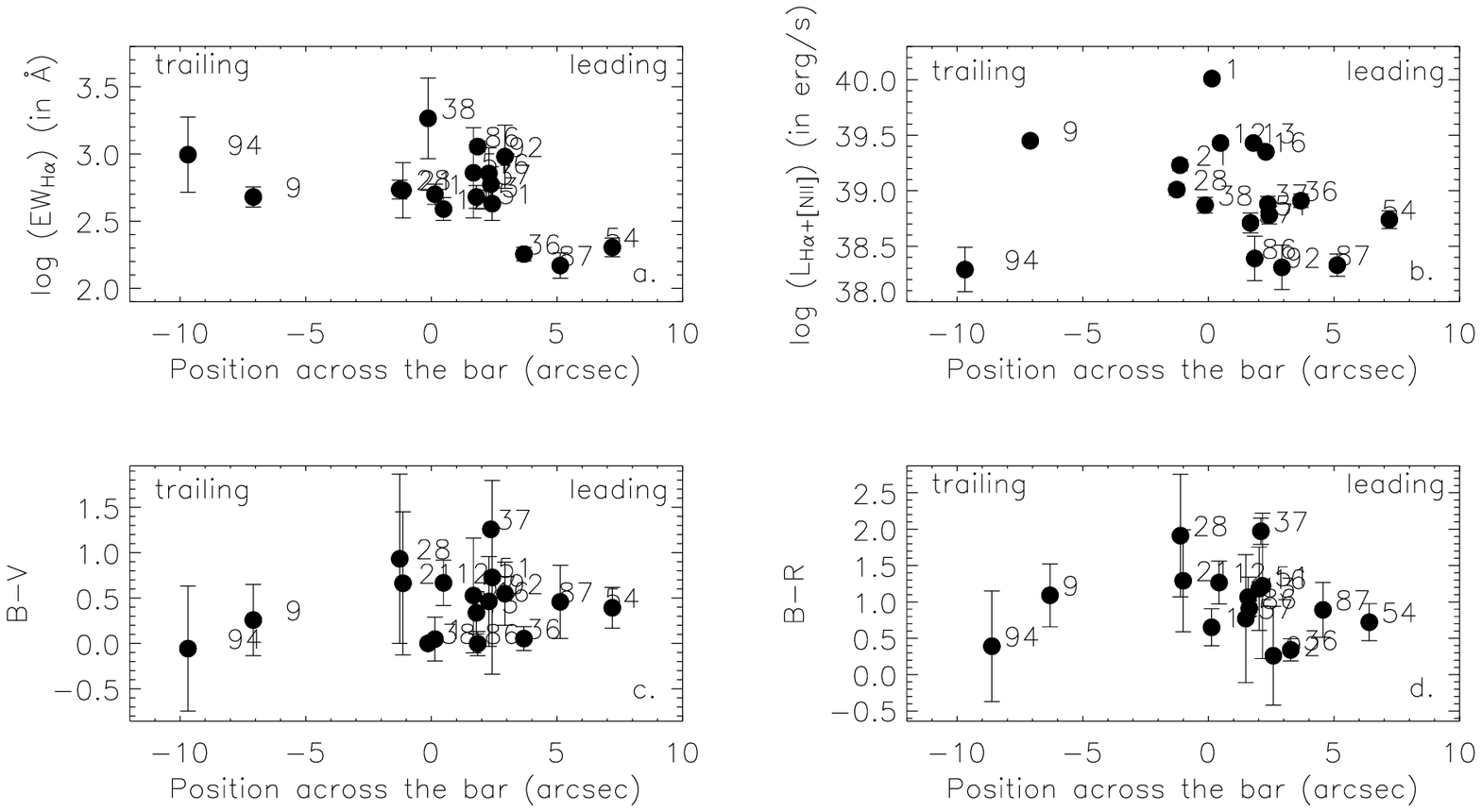}
\caption{Representation of (from top to bottom and from left to right) the \ha\ equivalent width 
(EW$_{\rm{H\alpha}}$), H$\alpha$+[\nii] luminosity, and $B-V$ and $B-R$ broad--band colours of 
the bar \hii\ regions  of NGC~1530 as a function of their deprojected 
distance from the bar dust--lane  in arc seconds. Zero means that the \hii\ region 
is located on the bar dust--lane, negative and positive values refer to \hii\ 
regions located on the trailing and leading side of 
the bar dust--lane respectively. The EW$_{\rm{H\alpha}}$ have been corrected for [\nii] 
contamination as described in Sect.~\ref{ha_imaging}. The broad--band colours have 
only been corrected for Galactic extinction (Schlegel et al. 1998).}
\label{resultsI}
%\end{figure*}
%\begin{figure*}[!h]
%\resizebox{\textwidth}
\includegraphics[width=1.0\textwidth]{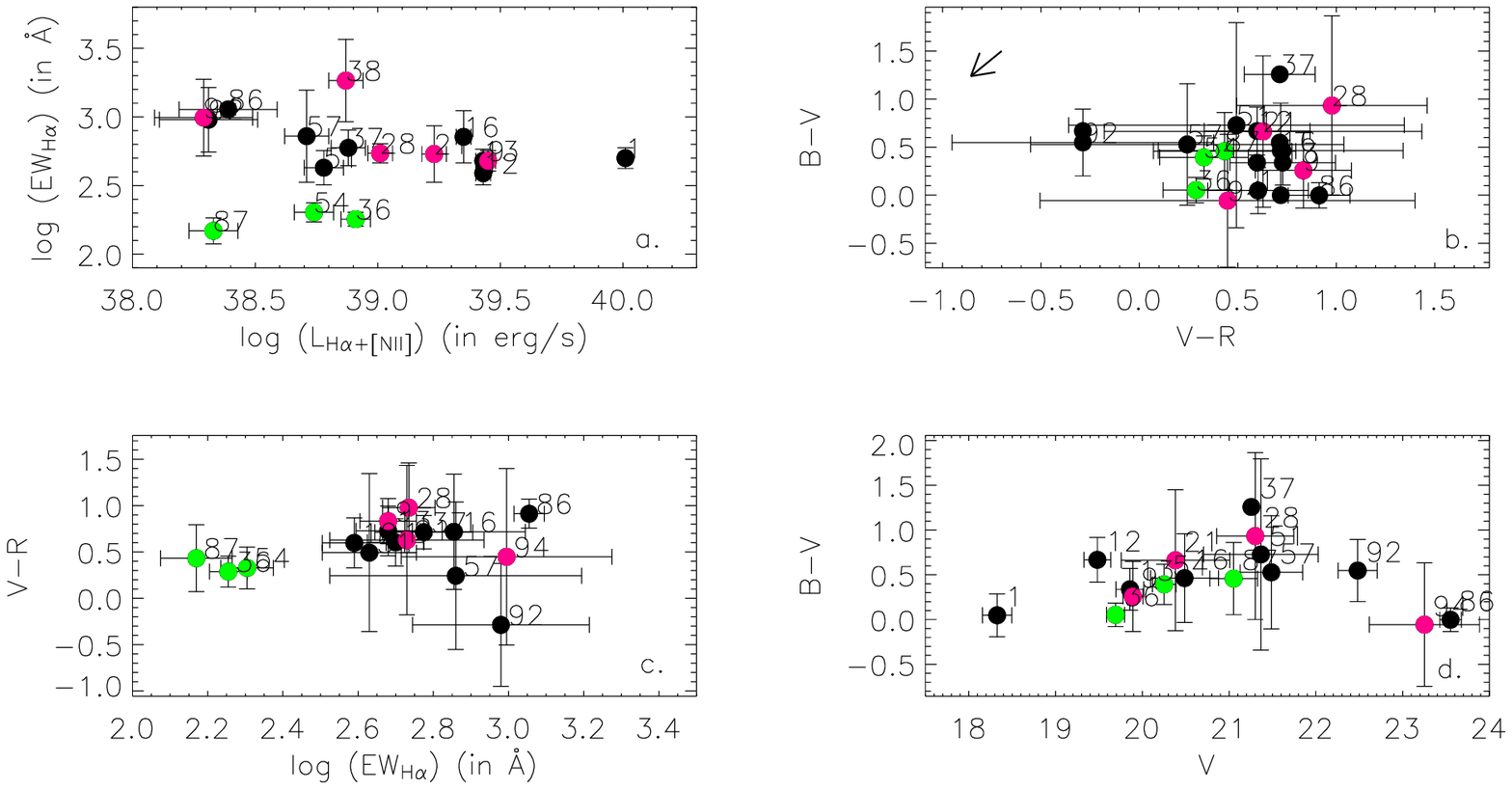}
\caption{Correlations between observed properties of the \hii\ regions of the bar of NGC~1530. 
Green dots correspond to the \hii\ regions with lowest observed \ha\ equivalent widths 
(located in the leading side of the bar dust--lane), pink dots correspond to \hii\ regions located 
in the trailing side of the bar dust--lane. The broad--band colours have only been corrected for 
Galactic extinction (Schlegel et al. 1998). The arrow in the $B-V$ vs. $V-R$ plot indicates the direction of the 
extinction correction corresponding to A$_V$=1 and assuming  Calzetti et al. (2000) attenuation curve.
The EW$_{\rm{H\alpha}}$ have been corrected for [\nii] contamination as described in 
Sect.~\ref{ha_imaging}. Note the lack of any systematic colour difference between \hii\ 
regions of the leading and trailing sides.}
\label{resultsII}
\end{figure*}

Figure~\ref{resultsI} shows the EW$_{\rm{H\alpha}}$, the \ha\ luminosity, and the
broad--band colours of the NGC~1530 bar \hii\ regions as a function 
of their deprojected distance to the bar dust--lane. 
We can see in Fig.~\ref{resultsI}a that the mean  $\log~($EW$_{\rm{H\alpha}})$ 
 is  $2.7\pm0.3$ (or EW$_{\rm{H\alpha}}\sim$660~\AA\ 
with a standard deviation of $\sim$450\AA). The measured  EW$_{\rm{H\alpha}}$ of the bar
\hii\ regions of NGC~1530 are within typical ranges for measurements of disc \hii\ regions
in spirals  (e.g. Bresolin \& Kennicutt, 1999; von Hippel \& Bothun,  1990; Cedr\'es et al. 2005). 

Our  EW$_{\rm{H\alpha}}$ measurements are, however, a bit higher than those measured by Martin \&
Friedli (1999) for bar \hii\ regions in several galaxies, 
which had a mean distribution value of $\sim$250\AA, but these authors 
do indicate that their measurements could be up to twice as large due to the 
contamination from galactic continuum in their measurements of stellar continuum.
 
It is interesting to note that the \hii\ regions 
located further from the bar dust--lane in the leading side of the bar have 
lower EW$_{\rm{H\alpha}}$, by a factor $\sim$4 -- 5, or a 0.6-0.7 dex 
difference in $\log$~EW$_{\rm{H\alpha}}$, than \hii\ regions located on 
the trailing side and those closer to the bar dust--lane (see 
Fig.~\ref{resultsI}a and Fig.~\ref{mosaico_barra}d). In terms of broad--band colours and \ha\ 
luminosities there is no obvious difference, within the margins of error, between 
the \hii\ regions of lower  EW$_{\rm{H\alpha}}$ and the remainder. 

We developed Monte Carlo simulations to measure the statistical significance 
of finding the lowest EW$_{\rm{H\alpha}}$ \hii\  regions in the leading side of 
the bar of NGC 1530;   was this due to chance 
or had some physical origin? We simulated a random, uniform distribution of 
EW$_{\rm{H\alpha}}$  values, at random positions across the bar, both leading 
and trailing, for 17 \hii\ regions, which were the number of regions for which we 
had reliable measurements of EW$_{\rm{H\alpha}}$.
%In order to test the statistical significance of having the 
%lowest EW$_{\rm{H\alpha}}$ \hii\ regions in the leading side 
%of the NGC~1530 bar, Monte Carlo simulations have been carried out.
%Is it simply due to chance or does it have a  physical origin? 
%We  simulated a random uniform distribution of EW$_{\rm{H\alpha}}$ values, 
%and random positions across the bar (leading or trailing) for a sample of 17 
%\hii\ regions (the number of regions of the bar of NGC~1530 for 
%which we have reliable measurements of EW$_{\rm{H\alpha}}$).
We then measured the position of the three \hii\ regions that had the lowest 
EW$_{\rm{H\alpha}}$. The test was repeated 
10$^{6}$ times for uniform and normal random position distributions. 
We  then measured the probability that the three \hii\  regions of the 
lowest measured EW$_{\rm{H\alpha}}$ values were located in the leading
side of the bar, by chance: we found that the likelihood was low and 
equal to approximately 12.5\%.
Therefore, it is very important to study the 
factors that can produce such differences in  EW$_{\rm{H\alpha}}$,
due to its potential information to elucidate the properties of SF 
in bar environments.

\section{Discussion on the measured \hii\ region EW$_{\rm{H\alpha}}$}

The most striking result from the analysis performed in the previous sections
comes from the lower measured EW$_{\rm{H\alpha}}$  for the \hii\ regions
located furthest away from the bar dust--lanes on the leading side of the bar.
The relevance of this result is enhanced by the fact that it links both 
recent star forming regions, through their  EW$_{\rm{H\alpha}}$, and the 
gas dynamics of the bar, through the bar dust--lanes. Understanding 
the origin of this relation can shed light on the processes triggering 
SF in bars.

The \ha\ equivalent width of \hii\ regions depends on various 
factors including: the evolutionary status 
of the \hii\ region, the initial mass function (IMF) of their associated 
stellar clusters, their metal content, the ionizing photon leakage from the 
region, or  the dust extinction of ionizing photons emitted from the stellar 
association. Contamination by [\nii] emission lines within the \ha\  filter 
bandwidth can also artificially increase the  EW$_{\rm{H\alpha}}$. Absorption 
by the underlying population of the bar/disc can, in principle, reduce the observed
EW; however, for the \ha\ line this effect is negligible, because the 
EW$_{\rm{H\alpha}}$ is $\sim2-5$\AA\ for all ranges of metallicities and ages 
(Gonz\'alez--Delgado et al. 2005).

In the next subsections, all of these factors will be discussed in detail. The aim
is to  constrain the agents which can differentially affect \hii\ regions 
of the bar  and therefore  be responsible of the lower observed
EW$_{\rm{H\alpha}}$ of the \hii\ regions of the leading side of the NGC~1530 bar.

\subsection{Observational effects: contamination by [\nii]}
\label{observational_effects}
[\nii] contamination in the  \ha\ fluxes of the \hii\ regions of the bar
can  affect differentially the measured  EW$_{\rm{H\alpha}}$, 
which we correct by assuming a constant [\nii]$\lambda$6584/\ha\ for all \hii\ 
regions (see Sect.~\ref{ha_imaging}). The [\nii]$\lambda$6584/\ha\  line 
ratio depends mainly on the metallicity and  ionization parameter, but 
 can also be enhanced due to strong shock excitation.

The observed [\nii]$\lambda$6584/\ha\ and [\sii]$\lambda$$\lambda$6717,6731/\ha\ 
line ratios for two bar  \hii\ regions and one circumnuclear region  in
NGC~1530 by M\'arquez et al. (2002),  are consistent with pure photoionization 
(Kewley et al. 2001), which excludes a strong contribution from shocks.
Spectroscopic observations of bar and disc \hii\ regions completed by 
Martin \& Friedli (1999)  showed that bar \hii\ regions do not exhibit any 
systematic evidence of shock ionization. In any case,  using the 
measured maximum deprojected non--circular velocities in the bar 
of NGC~1530 ($\sim$160 km~s\me, Zurita et al. 2004), we can estimate
 the [\nii] emission expected for shocks of this velocity. 
Dopita \& Sutherland (1996) models predict a maximum [\nii]$\lambda$6584/\ha\ line 
ratio ranging from $\sim$0.21 to $\sim$0.70 for a shock velocity of  
$\sim$200 km s\me\ and a magnetic--field parameter ($B/n^{1/2}$) varying 
between 0 and 4 $\mu G/cm^{3/2}$.

According to Fig.~\ref{contaminacion}a, this line ratio  implies a contamination 
from [\nii]$\lambda$6584+$\lambda$6548 in L$_{\rm{H\alpha}}$ and 
EW$_{\rm{H\alpha}}$ for our on--line \ha\ filter, ranging from $\sim$15\% to  
$\sim$38\% depending on  $B/n^{1/2}$.
Therefore, the maximum difference in  L$_{\rm{H\alpha}}$ and EW$_{\rm{H\alpha}}$ 
that we would expect to see from region to region if shock ionization were 
contributing differentially between \hii\ regions would be 
$\lesssim$38\% of the measured L$_{\rm{H\alpha}+[NII]}$ 
and  EW$_{\rm{H\alpha}+[NII]}$, or a 0.07--0.2~dex difference in 
$\log$~EW$_{\rm{H\alpha}+[NII]}$. 
This fails to explain the observed differences in 
$\log$~EW$_{\rm{H\alpha}}$ ($\sim$0.6--0.7dex) even 
for the extreme situation  in which the regions of lowest EW$_{\rm{H\alpha}}$   
would have a null contribution from shocks, whereas the remainder 
would have a maximum contribution from ionization produced by shocks.

The differential contamination due to differential metallicity
in the \hii\ regions of the bar can be quantified using 
Fig.~\ref{contaminacion}b, which was created using the 
Kewley \& Dopita (2002) models and the transmission curves of the two
filters used to obtain our  \ha\ observations. The figure shows 
the percentage of contamination on the measured L$_{\rm{H\alpha}}$, as a function of 
metallicity, for different values of the ionization parameter,
$q$, as defined by Kewley \& Dopita (2002). The values of $q$ were selected
to encompass the values observed for extragalactic \hii\ regions $10^7<q<10^8$ 
(Dopita et al. 2000).

To our knowledge there are no published metallicity measurements 
for the \hii\ regions of NGC~1530. Therefore we used M\'arquez 
et al. (2002) published line ratios (see Sect.~\ref{ha_imaging}), and 
the bright emission--line  diagnostic diagrams 
by Kewley \& Dopita (2002), to estimate a range of plausible 
metallicities for the bar \hii\ regions. Assuming as above that $10^7<q<10^8$ 
(Dopita et al. 2000), Figs. 4 and 7 of Kewley \& Dopita (2002) 
indicate a metallicity in the range 0.9Z$_\odot$-1.8Z$_\odot$
(or 8.6$\lesssim 12 + \log$ (O/H)$\lesssim  8.92$, assuming  
$12 + \log$ (O/H)$_\odot$ = 8.66, Allende--Prieto et al. 2001).

Figure 8 of Martin \& Friedli (1999) shows, for a sample of 10 
barred galaxies, that the maximum metallicity difference between 
\hii\ regions of the same galaxy bar  varies between $\sim$0.02 and 
0.18 dex in $12+\log(\rm{O/H})$. Therefore, if the mean metallicity of 
the NGC~1530 bar \hii\ regions is equal to  8.6 dex, which is our  
lower limit estimate, the  contamination due to changes in metallicity 
of $\pm0.09$~dex can vary between 0 and 17\% of 
the observed L$_{\rm{H\alpha}+[NII]}$. These values can increase 
 to $\sim$7--31\%, if the 
mean metallicity in the bar of NGC~1530 is 8.9 dex, which is the 
upper limit of our metallicity estimate.
Therefore, we could expect a maximum difference in $\log$~L$_{\rm{H\alpha}+[NII]}$, or 
$\log$~EW$_{\rm{H\alpha}+[NII]}$, between \hii\ regions, due to differential 
[\nii]--contamination produced by differences in metallicity, of up to $\sim$0.12 dex.

Therefore, differential [\nii] contamination on the measured L$_{\rm{H\alpha}+[NII]}$
fluxes, due to different shock contributions to ionization and/or changes in metallicity from 
region to region, can account for a maximum of a $\sim$0.32~dex difference in the measured 
$\log$~EW$_{\rm{H\alpha}}$, if:  (a) shock ionization affects only regions located 
on the trailing side, and closest to the bar dust--lanes, for a magnetic--field parameter 
of 4 $\mu G/cm^{3/2}$, and (b) only if these regions have also higher metallicity, 
by $\sim$0.18 dex, and lower $q$ for a mean bar metallicity of $\sim$8.9 dex. 
If the bar mean metallicity was lower and/or the conditions before were not met, 
the maximum difference in $\log$~EW$_{\rm{H\alpha}}$ that differential [\nii] 
contamination could account for, would be much lower than $\sim$0.32~dex.

\subsection{Ionizing photon extinction and leakage}
\label{polvo_escape}
One important factor that can produce a lower observed EW$_{\rm{H\alpha}}$ 
is the {\em loss} of ionizing photons before they can ionize the neutral 
hydrogen and produce \ha\ emission by subsequent electron 
recombination. This loss implies that the measured \ha\ luminosity accounts 
for only a percentage of the Lyman continuum photons (hereinafter 
Lyc) emitted by the OB association, and implies a lower than predicted
observed \ha\ equivalent width.

There is widespread  observational evidence that  non--negligible fractions
of Lyc photons are {\em lost} because of both dust absorption  
inside \hii\ regions (e.g. Inoue 2001), 
and the escape of Lyc photons from  the regions  (e.g. Zurita et al. 2000; 
Castellanos et al. 2002, Rela\~no et al. 2002). 
However, estimating the total fraction of Lyc photons {\em lost} from an 
specific \hii\ region is not straightforward because it requires knowledge 
of the spectral types for the complete census of individual stars that constitute
the ionizing cluster; we could therefore predict the emitted Lyc flux using 
stellar--atmosphere models and compare the predicted with the observed 
Lyc flux that is down--converted into recombination lines.
The determination of the missing fraction that can be attributed to dust 
extinction or photon leakage, is even more complicated.
Normally, no Lyc--flux dust--extinction is considered when escape fractions 
are being estimated, and vice versa.
Given the distance of NGC~1530,  estimating  whether dust or photon leakage 
could be the origin of the low EW$_{\rm{H\alpha}}$  of the regions on the leading 
side of the bar, is a complex task and we have to rely on empirical results 
based on statistical properties of \hii\ regions and/or theoretical models.

\begin{figure}[]
%\resizebox{\textwidth}
\includegraphics[width=0.5\textwidth]{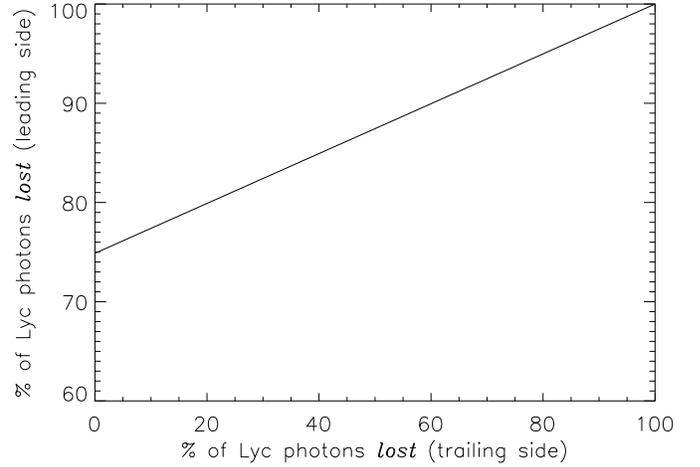}
\caption{Required percentage of Lyc photons absorbed by dust and/or escaping from the \hii\ regions 
of the leading side of bar dust--lane of NGC~1530, as a function of the same percentage for the \hii\ 
regions of the trailing side, under the hypothesis that ionizing--photon extinction
and photon leakage are the only causes of the observed difference in  EW$_{\rm{H\alpha}}$  
(Fig.~\ref{resultsI}a).}
\label{diff_lead_trail}
\end{figure}

Figure~\ref{diff_lead_trail} shows the required  percentage of {\em lost} Lyc photons
(due to dust extinction and/or leakage) from the \hii\ regions 
of lowest EW$_{\rm{H\alpha}}$, which are those located in the leading side, 
as a function of the same percentage for the remaining \hii\ regions. 
The underlying hypothesis is that ionizing--photon extinction
and photon leakage, are the only causes of the observed difference in  
EW$_{\rm{H\alpha}}$ (Fig.~\ref{resultsI}a). We now discuss whether current 
knowledge on these factors can give a natural framework for this plot.

Neither the \hii\ regions of lowest EW$_{\rm{H\alpha}}$, indicated by the 
green--filled dots in Fig.~\ref{resultsII}, nor their associated local backgrounds, 
indicated by purple--filled squares in Fig.~\ref{fondos_locales}, are particularly 
redder than the remainder of the regions. The \hii\ region colours 
depend on a number of factors, namely metallicity, age, and more importantly, 
the dust extinction. Disentangling these effects from the broad--band data 
alone is impossible. Unfortunately,  no Spitzer 
data is available for NGC~1530, which would  help  to constraint 
the dust extinction. 
%It appears that there is not much more dust extinction in the optical 
%bands, on the leading side of the bar, compared to the trailing side 
%of the bar, neglecting any geometrical dust--source distribution and 
%evolutionary effects, because, to first order, the \hii\ regions of 
%lowest EW$_{\rm{H\alpha}}$ are not the reddest studied.
However, to first order, the fact that the 
\hii\ regions of lowest EW$_{\rm{H\alpha}}$ are not the reddest of the sample
indicates, neglecting any geometrical dust--source 
distribution and evolutionary effects, that there is
not much more dust extinction in the optical bands in the leading
than in  the trailing side of the bar. 
This implies  that a similar result may be obtained for 
the ionizing--flux extinction.

Dopita et al. (2003) parametrised the percentage of Lyc photons absorbed by 
a mixture of dust grains distributed in a spherical shell inside an \hii\ region. 
This percentage  increases as  the metallicity,  ionization parameter, and the 
content of complex polycyclic aromatic carbon  (PAH) molecules increases. 
Assuming that metallicity is approximately constant across the bar of NGC~1530,
and assuming an ionization parameter in the range $10^7<q<10^8$  (or $-3.5<\log U<-2.5$; 
Dopita et al. 2000), the parameter combination that would account for the largest 
difference between the leading and trailing sides, for a given metallicity,
would imply  that \hii\ regions 
on the trailing side have the lowest $q$ ($10^7$) and 0\% content of PAHs in their 
dust, while \hii\ regions on the leading side would have $q=10^8$ and 20\% of PAHs,
which is an observational upper limit for these molecules. Even in such 
extreme situations, there would be a $\sim$1\% and  $\sim$28\%  
absorption of the total emitted Lyc photons in the trailing and leading 
\hii\ regions respectively, assuming a  metallicity of  
$12 + \log$~(O/H)$\sim$8.6. These quantities would increase by up to  $\sim$3\%
and $\sim$42\% in the  trailing and leading sides for
 $12 + \log$~(O/H)$\sim$8.92. 
In any case, and if this spherical shell dust morphology were representative 
of the \hii\ regions in the bar of NGC~1530,
the dust extinction could not explain completely observed differences 
in EW$_{\rm{H\alpha}}$  because $\sim30\%$ additional extinction on the 
leading side would be required, as can be seen in Fig.~\ref{diff_lead_trail}.

A alternative method to estimate the dust extinction of Lyc photons, 
considers an inhomogeneous interstellar--medium in which gas and dust 
are distributed in optically--thick clumps of high--density, which are 
surrounded by a component of lower density (Giammanco et al. 2004; 
Beckman \& Guti\' errez, private communication). In this scenario, 
the percentage of Lyc photons absorbed by dust is determined by 
the degree of ionization of the clumps. This translates into a 
dependence of the Lyc extinction, on the size and the geometrical filling 
factor\footnote{The geometrical filling factor is defined to be the 
fraction of the total volume of the  \hii\ region occupied by the clumps. It is determined 
by the clump size and the number density of clumps inside the \hii\ region (Giammanco et al. 2004).} 
of the \hii\ region, for a given extinction coefficient or dust grains properties.
The \hii\ regions of lowest  EW$_{\rm{H\alpha}}$  are not particularly different in 
terms of size than the rest (see e.g. Fig.~\ref{mosaico_barra}a).
Assuming the same extinction coefficient for both sides of the bar,
the geometrical filling factor would need to be approximately seventy
times larger in the \hii\ regions of the leading side with respect 
to that of the trailing side, if the {\em lost} fraction 
in the trailing side were close to zero, and therefore $\sim75$\% in the leading side, 
(see Fig.~\ref{diff_lead_trail}). A geometrical filling factor 
that is more than three times larger would be required to explain 
the {\em lost} fractions that are as high as  $\sim$60\% and 
$\sim$90\%, for the trailing and leading sides, respectively.
%A $\sim3$ times larger geometrical filling 
%factor would be required for {\em lost} fractions as high as  $\sim$60\% and 
%$\sim$90\% in the trailing and leading sides respectively. 
The same scaling 
relations apply to the extinction coefficient if the geometrical filling 
factor is assumed to be constant (Beckman \& Guti\' errez, private communication).

Therefore,  to explain the differences between the regions of the 
leading and trailing sides of the bar of NGC~1530 in terms of dust extinction alone, 
important differences in the  interstellar--medium properties, such as the degree of
clumpiness and/or composition of the dust, would be required. 
For dust--shell morphology, dust extinction alone cannot explain the 
differences observed in EW$_{\rm{H\alpha}}$; for clumpy models, the required change 
in filling factor on such small scale  appears unfeasible. In any case, 
our optical data do not provide clear evidence of enhanced extinction in 
the regions with the lowest measured EW$_{\rm{H\alpha}}$.
%In the case of
%the dust shell morphology, dust extinction alone falls short to explain the 
%observed differences in EW$_{\rm{H\alpha}}$ and in the case of clumpy models, 
%the required change in filling factor in such a small scale seems unfeasible, 
%and in any case, our optical data do not show any clear 
%evidence of enhanced extinction for the regions of lowest EW$_{\rm{H\alpha}}$.

Ionizing--photon escape fractions have been directly or indirectly  measured 
to range from 0\%  up to even $\sim$80\% of the total Lyc luminosity emitted by
the corresponding OB association (e.g. Zurita et al. 2000, 2002; 
Castellanos et al. 2002, Rela\~no et al. 2002; Oey \& Kennicutt 1997). 
The luminosity functions of \hii\ regions in nearby spirals and the distribution 
of diffuse ionized gas in NGC~157, suggest that the escape fraction increases 
as the observed L$_{\rm{H\alpha}}$ increases (Zurita et al. 2000, 2002; Beckman 
et al. 2000). Figure~\ref{resultsII}a, however, indicates that the measurements 
of L$_{\rm{H\alpha}}$ for the regions with the lowest measured EW$_{\rm{H\alpha}}$, 
are not amongst the highest values measured.
%However, Fig.~\ref{resultsII}a shows that the  L$_{\rm{H\alpha}}$ of 
%the regions of lowest  EW$_{\rm{H\alpha}}$ are not  within the highest of the sample.

The escape of ionizing photons from \hii\ regions appears to be a
natural consequence of the inhomogeneity  of the interstellar medium
(Giammanco et al. 2004). For this model the fraction of photon 
leakage depends on the \hii\ region radius and on the geometrical 
filling factor described above.
The leakage fraction increases as the \hii\ region radius 
and/or the geometrical filling factor decrease  (see Fig. 2 of Giammanco et al. 2004).
As mentioned above, the  \hii\ regions of the leading side do not differ in size 
from  the remainder regions. 
Using the clumpy model, the \hii\ regions located further away 
from the bar dust--lane on the leading side, would need to have 
smaller geometrical filling factors, or smaller clumps for a fixed \hii\ 
region size, than those located on the trailing side, to allow a higher 
fraction of ionizing photons to leak.
%Under the clumpy model, in order to allow for a higher fraction of 
%ionizing photons to leak, the \hii\ regions located further away from
%the bar dust--lane in the leading side, would need to have 
%smaller geometrical filling factor (or smaller clumps for a fixed \hii\ 
%region size) than the ones located in the trailing side.
Assuming similar clump size, the fraction of the volume of an \hii\ region 
occupied by clumps (i.e. the geometrical filling factor) 
would be required to be a least 10 times less in the leading side 
(Giammanco et al. 2004) to produce the leakage implied
by Fig.~\ref{diff_lead_trail}.

\subsection{Metallicity,  IMF and age}
\label{starburst}
The  EW$_{\rm{H\alpha}}$ is a measure of the relative amount of ionizing 
and continuum photons emitted by the whole stellar cluster associated with
the \hii\ region. It  depends on the evolutionary status of the 
stars, the initial mass function (IMF), and  metallicity. 
Evolutionary--synthesis models such as {\em Starburst99} (Leitherer 
et al. 1999) can be used to estimate the  influence of each parameter on 
the EW$_{\rm{H\alpha}}$. The NGC~1530 bar \hii\ region 
metallicity, which was estimated as described in Sect.~\ref{observational_effects},
was used as an input for our {\em Starburst99} simulations.

\begin{figure}[!h]
%\resizebox{\textwidth}
\includegraphics[width=0.5\textwidth]{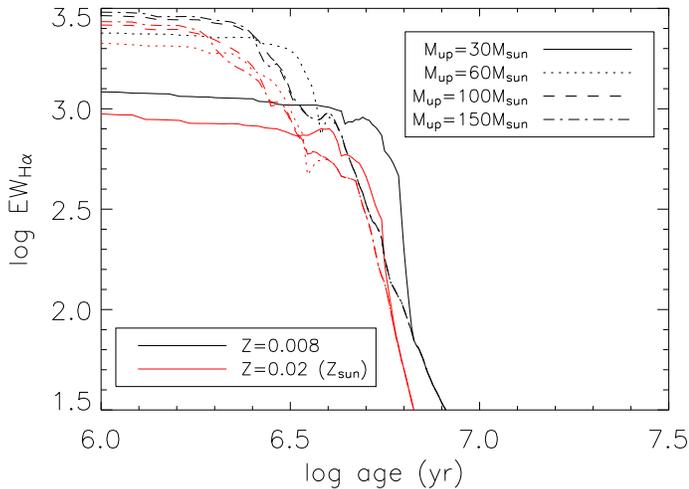}
\caption{Evolution of the \ha\ equivalent width as a function of the age  predicted 
by theoretical models (Leitherer et al. 1999), for an IMF with exponents  1.3 and 
2.35, for masses below and above 0.5M$_\odot$, respectively. Different IMF upper 
mass limits were assumed: 
30M$_\odot$ (solid lines), 60M$_\odot$ (dotted lines), 100M$_\odot$ (dashed lines), 
150M$_\odot$ (dot--dash--dot lines). Red and black lines indicate metallicities of Z=0.008 and 
Z=0.02 (solar), respectively.}
\label{ew_age}
\end{figure}

Figure~\ref{ew_age} shows the decimal logarithm of the modelled  
EW$_{\rm{H\alpha}}$ as a function of the decimal logarithm 
of the \hii\ region age for different  simulations. We assume a 
multi--power--law parametrisation  of the IMF 
($dN/dm \propto m^{-\alpha}$) with exponents $\alpha$=1.3 
for $0.1{\rm M}_\odot<{\rm M}<0.5{\rm M}_\odot$ and $\alpha$=2.35 (Salpeter) for 
${\rm M}>0.5{\rm M}_\odot$. Each simulation corresponds to a different 
upper mass (M$_{up}$) boundary for the IMF 
(30, 60, 100 or 150M$_\odot$) for two  metallicities, 
Z=0.02 (solar) and Z=0.008\footnote{The Geneva stellar evolutionary 
tracks employed by {\em Starburst99} are still on the scale of the 
old solar abundance (Vazquez \& Leitherer, 2005), in which 
$12 + \log$ (O/H)$_\odot$=8.9. On that scale, the metallicity of NGC~1530 is 
0.5Z$_\odot$-Z$_\odot$, which determines our selection of metallicity 
for the {\em Starburst99} simulations.}, which includes the range of 
estimated metallicity of the bar of the galaxy.

Differences in
the IMF have a larger impact  in $\log$~EW$_{\rm{H\alpha}}$ for ages
$\lesssim2.5$~Myr (e.g. Bresolin \& Kennicutt, 1999). 
Afterwards, the  EW$_{\rm{H\alpha}}$ becomes almost 
degenerate with the IMF;  the age of an \hii\ region can be determined
approximately when both its EW$_{\rm{H\alpha}}$ and metallicity are known.
We  employed the up--to--date version of {\em Starburst99}, v5.1, 
which uses the Geneva stellar--evolutionary tracks with high mass--loss 
(Schmutz et al. 1992). This version uses non--LTE line--blanketed 
model atmospheres (Smith et al. 2002), which produce a 
larger dependence on metallicity in the relation 
$\log$~EW$_{\rm{H\alpha}}$--$\log$~age than in the original 
Leitherer et al. (1999) models.

\begin{figure}[!h]
%\resizebox{\textwidth}
\includegraphics[width=0.5\textwidth]{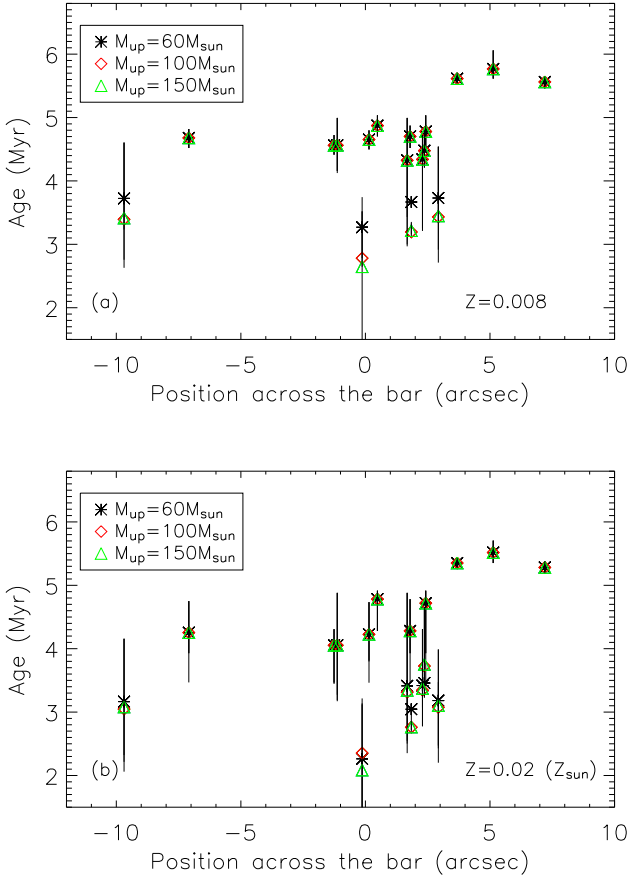}
\caption{Estimated age of the \hii\ regions of the bar of NGC~1530 as a function of their 
deprojected distance across the bar of the galaxy. Zero means that the \hii\ region is 
located on the bar dust--lane. Negative and positive values refer to \hii\ regions located 
on the trailing and leading sides of the bar dust--lanes respectively. Asterisks, diamonds 
and triangles indicate ages estimated from IMFs with upper mass limits of 60M$_\odot$, 
100M$_\odot$ and  150M$_\odot$, respectively. Top and bottom figures show ages when a 
metallicities of Z=0.008 and Z=0.02 (solar) are assumed, respectively. The error bars 
come from the conversion into age of the upper and lower bounds to the measured  
$\log$~EW$_{\rm{H\alpha}}$.}
\label{age_distance}
\end{figure}

Assuming that the \hii\ regions of the bar of NGC~1530 are the result 
of a single burst of SF, and neglecting other factors 
that affect the EW$_{\rm{H\alpha}}$, which were described 
in previous subsections, we can use Fig.~\ref{ew_age} to 
determine the age of each \hii\ region. This age is the
average value  for all  unresolved, non--coeval, star--clusters,
which is weighted by the age and size of each cluster, that 
probably coexist inside each giant \hii\ region. 
Figure~\ref{ew_age} shows 
that the evolutionary tracks associated with M$_{up}$=30M$_\odot$ 
are unable to explain the highest observed EW$_{\rm{H\alpha}}$,
and therefore will not be considered  hereinafter.
%(e.g. Bresolin \& Kennicutt 1999; **REFS HST images)
The  calculated \hii\ region ages, as a function of their 
deprojected distance to the bar dust--lane, are plotted 
in Figs.~\ref{age_distance}a and \ref{age_distance}b, for
metallicities of Z=0.008  and  Z=0.02 (solar), respectively.
The error bars represent the translation into age of 
the upper and lower bounds to the measured  
$\log$~EW$_{\rm{H\alpha}}$ (see Table~\ref{results_HII}).

As expected from Fig.~\ref{ew_age}, larger differences in age
estimates from different  M$_{up}$ occur for the 
regions with  the highest values of EW$_{\rm{H\alpha}}$, that is the
youngest regions.  The dispersion in ages 
for these high--EW$_{\rm{H\alpha}}$  \hii\ regions also
increases with metallicity. Figure~\ref{age_distance} indicates
that the \hii\ regions located further away from the bar dust--lane
on the leading side, are, on average, from $\sim$1.4 Myr to 1.7 Myr older 
than the remaining regions, depending on whether Z=0.008 or Z=0.02, respectively.

We should take into account that this age difference, in the 
absence of other factors affecting differentially both sets 
of regions,  must be considered as a lower bound. 
As discussed in Sect.~\ref{polvo_escape}, a 
percentage of ionizing photons are absorbed by dust inside
the \hii\ regions and/or escape from the region. Assuming a conservative
value of $\sim$50\% of ionizing photons {\em lost} in all \hii\ 
regions of the bar, and repeating the exercise of 
Fig.~\ref{age_distance}, we derive mean differences ranging 
from $\sim$1.5 to $\sim$2.5Myr, for the ranges of M$_{up}$ 
and metallicity considered in the {\em Starburst99} simulations.

In summary, there are a number of factors that can 
decrease  the EW$_{\rm{H\alpha}}$ of
the \hii\ regions which are located furthest away from the NGC~1530 
bar dust--lane. Of these, the only factor that could explain 
the EW$_{\rm{H\alpha}}$ measurements naturally  is age. 
Other effects  such as Lyc--photon dust--extinction and photon leakage, 
could, in principle, in combination  explain the observations, but 
they would imply important differences in dust composition 
and interstellar--medium structure within the bar region, on 
scales of $\sim$1~kpc, that appear infeasible. Therefore, it is
natural  to wonder why the \hii\ regions of the 
leading side are  older than the remainder.
This question will be discussed in then next section.

\section{Discussion}
\label{discuss_older}

We find evidence that the \hii\ regions on the leading side 
of the bar dust--lane of NGC~1530 are on average $\sim$1.5-2.5 Myr older,
than  regions  on the trailing side of the bar dust--lane. This
 implies that  there has been either a time delay and a location
offset for the star formation bursts in the bar of NGC~1530, or 
 an ageing and migration effect of the stars formed initially 
on the trailing side of the shock region of the bar, towards the leading side.

A priori, there is no indication to suggest that the SF was favoured 
first on one side of the bar dust--lane (the leading side) and later on, 
the SF burst has shifted towards the trailing side of the dust--lane. This would imply 
that on timescales shorter than a few Myrs, the conditions triggering SF 
have changed, to favour the trailing side with respect to the leading side 
of the bar dust--lane. It appears infeasible that the bar dynamics and matter
distribution have changed in such an insignificant time when compared to the 
bar orbital--time.

To determine whether the age difference is indeed due to migration 
from the trailing to the leading side of the bar dust--lane, we develop 
a test based on the timescale, that is a comparison between the mean 
age difference of \hii\ regions located in both sides of the bar dust--lane, 
and the dynamical crossing--time of a disc cloud/star, in the frame 
that rotates with the bar.
Taking into account the radial galactocentric 
distance for the lowest EW$_{\rm{H\alpha}}$ \hii\ regions, their 
corresponding disc rotational velocity (Zurita et al. 2004), and  
the pattern speed of the bar (10~km~s\me~kpc\me, 
P\'erez \& Zurita 2008), we  expect a
$\sim$0.5~kpc separation of the \hii\ regions from the bar dust
lane in $\sim$2 Myr. The measured distances from the bar dust--lane range  between
$\sim$0.6--1.2~kpc. Given the  approximation of this estimate we
cannot extract firm conclusions, but it shows that the timescales are
dynamically plausible to support the migration hypothesis.

One important factor to take into account in this migration 
hypothesis regards the survival of the star--forming clouds 
entering the dust--lanes of bars. This effect has been studied 
by Tubbs (1982). He tested the destruction of
dense, pre--stellar clouds in the gravitational field of a strong bar
(NGC~5383), and showed that clouds that encounter the straight dust
lane at velocities higher than 20 to 60~km~s\me\ must disperse. The cloud
collapse is  believed to be caused by an increase in the
stellar pressure on the pre--stellar cloud. 
This pressure increase is directly proportional to the 
density increase, for a constant temperature. 
Therefore, not only the relative velocity, but  the
fractional density enhancement  influences the collapse and 
survival of the cloud. Parameters outside a narrow region of density
enhancement and relative velocity  do not inhibit the star
formation along the bar (Tubbs 1982). For higher relative velocity, larger
fractional density enhancement is required  inhibit the
pre--stellar cloud survival.  For NGC~1530, the conditions for 
pre--stellar cloud survival must be fulfilled because star 
formation proceeds along the bar; we discuss this point further below.
%since there 
%is star formation along the bar, the conditions for pre--stellar cloud 
%survival must be fulfilled (this point will be further discussed below).

The hypothesis that stars could form in the trailing side of the 
bar dust--lane and then migrate towards the leading side, was proposed by 
Sheth et al. (2000).  They hypothesized that stars form 
in dust spurs, or observed dust--lanes which are approximately 
perpendicular to the bar, on the bar trailing side. They then  
continue their elliptical orbits until they ballistically pass through 
the shocks, inside the main bar dust--lanes, and ionize the neutral
gas at the leading side of the bar. Their proposal results 
from the observed spatial correlation between dust spurs 
seen in a $R-K'$ colour map and the location of the \hii\ regions 
along the bar of NGC~5383. Further correlation was found in 
a later work, by the presence of faint CO spurs towards the
trailing side of the main bar dust--lane (Sheth et al. 2002)
of a sample of spirals.
Other authors have also noticed the presence of dust spurs 
associated with galaxies with current star formation along the bars
(Martin \& Friedli 1997).

In addition to the spatial correlation between dust spurs and the location
of the \hii\ regions of the bar of  NGC~1530 (Sect.~\ref{morfo}), 
the fact that there are several \hii\ regions located in the trailing 
side of the bar of NGC~1530, and that these regions are younger than
the ones of the leading side, provides further support to the
Sheth et al. (2000, 2002) hypothesis. Our results are compatible
with the ageing of the recently--formed stars in the spurs as they cross the 
bar in a timescale compatible with the bar dynamics. 

Further evidence supporting this hypothesis comes from 
the \ha\ kinematics. A previous study of the kinematics 
of the ionized gas of NGC~1530, revealed the presence 
of velocity gradients in the direction parallel to 
the bar, as visible in Fig.~11c of Zurita 
et al. 2004, with strength $\sim0.13-0.20$~km~s\me~pc\me, which is 
lower than those associated with the main bar dust--lane.
The sites of larger velocity gradients in the direction of 
the bar, define lines approximately perpendicular to the 
main bar dust--lane. An offset between these velocity gradients and
the centroids of the \hii\ regions of the bar was already 
noticed by Zurita et al. (2004), but no conclusion about their 
origin was reached.
The comparison of the velocity gradients parallel to the bar
and our colour maps of NGC~1530,  shows very  good spatial 
agreement, most notably in the NW side, between the location of 
dust spurs and the loci of maximum velocity gradient (see 
Fig.~\ref{mosaico_barra}f).
This strongly suggests that the 
gradients observed are tracing flows of gas along the spurs towards 
the main bar dust--lane, of approximately constant velocity. If confirmed, 
this flow direction along the dust spurs would indicate that in a reference 
frame comoving with the bar, the gas flowing along the dust spurs 
would feed the main bar dust--lane.
This relative velocity gas--bar is lower in the 
spurs than in regions outside them, by $\sim25-50$~km~s\me, 
as can be seen from the dips in the residual velocity map 
(Fig. 11e of Zurita et al. 2004). Although we cannot discard 
a contribution from outflows from individual  \hii\ 
regions on the observed velocity gradients 
(Rela\~no et al. 2005),  this cannot 
be the dominant effect. The reason is that 
some of the brightest \hii\ regions 
in the arms do not show obvious associated gradients, and in 
any case, this outflow should be symmetric, and in some cases is 
only observed towards one side of the \hii\ regions, even for
regions located in the bar.

Dust spurs are also observed in the context of spiral--arm structure. These
dust spurs run perpendicularly towards the leading side of the main
dust--lane, the one that draws the spiral arm. In the frame of reference 
rotating with the spiral pattern, the gas arrives from the leading side of the arm 
to the main dust--lane, in contrast to the bar, for which
in its frame of reference, the gas would reach the
shock from the trailing side of the bar. Several hypotheses have been proposed to
explain the formation of these dust spurs: magnetic
effects (Kim \& Ostriker 2002), and gravitational instabilities (Chakrabarti et al. 
2003; Wada \& Koda 2004; Dobbs \& Bonnell 2006). There are observations
suggesting that star formation occurs in the dust spurs, and then
flows  towards the main dust--lane (La Vigne et al. 2006).
Therefore, it is reasonable to believe that the same phenomenon could be
occurring in the dust spurs of the bar region, because of the
clear association between the \hii\ regions and spurs already mentioned.

In summary,  the measured \hii\ region age--differences
across the bar are consistent with star formation occurring in 
the trailing side of the main bar dust--lanes, possibly even in the dust spurs,
and continuing their orbits towards the leading side of the main 
bar dust--lane. There is a theoretical framework for this hypothesis in
the context of the spiral arms.

To our knowledge, this paper contains the first measurement of 
differences in \hii\ region ages in a bar environment, which are directly 
linked to  dynamical parameters of a bar, by the \hii\ region 
positions relative to the main bar dust--lanes. As summarised in 
Sect.~\ref{intro}, so far most effort to understand  SF in bars 
has concentrated on the general morphology of the \ha\ emission and the spatial 
relation between the different stellar and gaseous components 
(e.g. Martin \& Friedli 1997; Verley et al. 2007b; Sheth et al. 2000, 2002).
We show here that not only the general distribution of the 
 \ha\ emission in the bar, but the detailed 
positions and properties of the \hii\ regions, with respect to the small--scale
dynamical and morphological features of the bar, are necessary to understand 
the conditions under which stars form.
We believe that the results presented here, open a new way to study the 
values of the parameters that cause SF to occur in bar environments and provide 
valuable information to constrain dynamical models willing 
to predict  star formation in bars.

\section{Summary and conclusions}
We have carefully analysed the photometric properties 
of the bar of NGC~1530 and  its \hii\ regions, which 
are studied in the context of their spatial relation with bar dynamical features.
Our main conclusions are:

\begin{itemize}

\item  The bar of NGC~1530 has flat surface--brightness profiles 
       along the bar in all bands ($B,V,R,Ks$).
       The broad--band ($B,V,R,Ks$) surface--brightness 
       profiles across the bar are rather 
       asymmetric and vary  with colour: the $B,V$ 
       and $R$ band profiles are steeper in the trailing 
       side than in the leading side, while this trend 
       is opposite in the $K$ band.

\item  The \ha\ equivalent--widths and broad--band colours 
       of the \hii\ regions of the bar have been computed 
       after a careful measurement of the local--background emission. 
       The local--background colours are compatible with those
       expected for the underlying  bar/disc stellar population
       when a colour--excess  E($B-V$)=0.25-0.35 is assumed.

\item  There is no obvious difference in terms of colours,  
       size or \ha\ emission, between the \hii\ regions 
       located in the trailing side of the bar dust--lane 
       and those located in the leading side.

\item  The EW$_{\rm{H\alpha}}$ of the \hii\ regions 
       located furthest away from the bar dust--lane, 
       on its leading side, is lower than the 
       EW$_{\rm{H\alpha}}$ of the rest of the \hii\ regions
       by a factor $\sim 4 - 5$ (or $\sim0.6-0.7$ dex difference 
       in $\log$~EW$_{\rm{H\alpha}}$).

\item  We have analysed the factors that could account 
       for the observed difference in EW$_{\rm{H\alpha}}$: [\nii] 
       contamination,
       Lyc--photon dust--extinction, the escape of Lyc photons, differences 
       in metallicity, IMF and age. Of these, only
       the age  can account for the observed differences.
       The measured difference in  EW$_{\rm{H\alpha}}$ implies that 
       the  \hii\ regions located furthest away from the bar 
       dust--lane, in its leading side, are on average 
       $1.5-2.5$~Myr older than the rest.

\item  Colour maps of NGC~1530 show the presence of dust spurs 
       on the trailing side of the bar main dust--lane. The  
       dust spurs reach the main bar dust--lane approximately 
       perpendicularly, and there is a clear spatial correlation 
       between the spurs and the position of \hii\ regions of the bar.

\item  There is a good spatial correlation  between the    
       location of the spurs and the zones of high velocity gradient 
       in the direction of the bar, strongly suggesting that the 
       gradients are tracing gas flow  along the spurs towards 
       the main bar dust--lane. This implies that dust spurs are zones 
       of higher gas density and lower relative velocity with 
       respect to the bar than the surrounding regions.
        
\item  The older age for the \hii\ regions of the leading side, together 
       with the spatial correlation with dust spurs and velocity gradients,
       provides further support to the hypothesis that star formation in bars 
       occurs in spurs. The ageing of the recently--formed stars, as
       they cross the bar, is compatible with the bar dynamics timescale.

\end{itemize}

We believe that the present work shows new important results 
which  favour the hypothesis that star formation in bars takes place
principally on the trailing side of the bar, in places where 
high gas density is present, although not the highest of the bar,
but simultaneously  low shear and low relative velocity 
with respect to the bar is found. This  highlights the importance of 
taking into account the dynamics of the bar. In addition, 
our findings are in good agreement with previous modelling and 
observations of star formation in the spiral arms.

A detailed study of  the bar \hii\ regions for a larger sample 
of galaxies,  to study their distribution with respect to 
the dynamical features of the bar, is a necessary step 
to confirm whether the findings of this paper can be generalized to
other bars with SF, and therefore constrain the parameters 
that allow SF to occur.

\begin{acknowledgements}
We are very grateful to Pierre Martin, John Beckman and M\'onica Rela\~no  for 
useful comments on the manuscript and Pablo P\'erez-Gonz\'alez 
for useful discussions on local background subtraction.
We acknowledge J.H. Knapen for kindly providing the $Ks$--band 
data of NGC~1530. We would like to thank the anonymous referee for comments.
A. Zurita acknowledges support from the Consejer\'\i a de 
Eduaci\'on y Ciencia de la Junta de Andaluc\'\i a. I. P\'erez   
is supported by a postdoctoral fellowship from the Netherlands Organization for 
Scientific Research (NWO, VENI-Grant 639.041.511) and 
the Spanish Plan Nacional del Espacio del Ministerio de Educaci\'on y Ciencia.
The WHT is operated on the island of La Palma by the Isaac 
Newton Group in the Spanish Observatorio del Roque de los Muchachos of the 
Instituto de Astrof\'\i sica de Canarias. 
Based on observations made with the Nordic Optical Telescope, operated
on the island of La Palma jointly by Denmark, Finland, Iceland,
Norway, and Sweden, in the Spanish Observatorio del Roque de los
Muchachos of the Instituto de Astrof\'\i sica de Canarias.  
\end{acknowledgements}

\end{document}